\documentclass[journal]{IEEEtran}
\ifCLASSINFOpdf
  \usepackage[pdftex]{graphicx}
\else
\fi
\usepackage{amsmath}
\usepackage{amsfonts}

\usepackage{algorithmic}
\usepackage{algorithm}

\usepackage{multirow}

\usepackage{url}
\begin{document}
\title{A Simple Model for Subject Behavior \\
in Subjective Experiments}

%\author{Michael~Shell,~\IEEEmembership{Member,~IEEE,}
%        John~Doe,~\IEEEmembership{Fellow,~OSA,}
%        and~Jane~Doe,~\IEEEmembership{Life~Fellow,~IEEE}% <-this % stops a space
%\thanks{M. Shell was with the Department
%of Electrical and Computer Engineering, Georgia Institute of Technology, Atlanta,

\author{Zhi~Li,
        Christos~G.~Bampis,
        Lukáš~Krasula (Netflix) \\
        Lucjan~Janowski (AGH, Poland) \\
        Ioannis~Katsavounidis (Facebook)}

\maketitle

\begin{abstract}
In a subjective experiment to evaluate the perceptual audiovisual
quality of multimedia and television services, raw opinion scores
collected from test subjects are often noisy and unreliable. To produce
the final mean opinion scores (MOS), recommendations such as ITU-R
BT.500, ITU-T P.910 and ITU-T P.913 standardize post-test screening
procedures to clean up the raw opinion scores, using techniques such
as subject outlier rejection and bias removal. In this paper, we analyze
the prior standardized techniques to demonstrate their weaknesses.
As an alternative, we propose a simple model to account for two of
the most dominant behaviors of subject inaccuracy: bias and inconsistency.
We further show that this model can also effectively deal with inattentive
subjects that give random scores. We propose to use maximum likelihood
estimation to jointly solve the model parameters, and present two
numeric solvers: the first based on the Newton-Raphson method, and
the second based on an alternating projection (AP). We show that the
AP solver generalizes the ITU-T P.913 post-test screening procedure
by weighing a subject's contribution to the true quality score by
her consistency (thus, the quality scores estimated can be interpreted
as bias-subtracted consistency-weighted MOS). We compare the proposed
methods with the standardized techniques using real datasets and synthetic
simulations, and demonstrate that the proposed methods are the most
valuable when the test conditions are challenging (for example, crowdsourcing
and cross-lab studies), offering advantages such as better model-data
fit, tighter confidence intervals, better robustness against subject
outliers, the absence of hard coded parameters and thresholds, and
auxiliary information on test subjects. The code for this work is
open-sourced at https://github.com/Netflix/sureal.
\end{abstract}

%\begin{IEEEkeywords}
%IEEE, IEEEtran, journal, \LaTeX, paper, template.
%\end{IEEEkeywords}

% \fi
\IEEEpeerreviewmaketitle

\section{Introduction}
\IEEEPARstart{S}{ubjective} experiment methodologies to evaluate the perceptual audiovisual
quality of multimedia and television services have been well studied.
Recommendations such as ITU-R BT.500 \cite{bt500}, ITU-T P.910 \cite{p910}
and ITU-T P.913 \cite{p913} standardize the procedures of conducting
subjective experiments and post-processing raw opinion scores to produce
the mean opinion scores (MOS) of test stimuli (e.g., a set of encoded
videos). To account for the inherently noisy and often unreliable
nature of test subjects, the recommendations have included corrective
mechanisms such as subject rejection (BT.500, and also referenced
in P.910 and P.913), subject bias removal (P.913), and criteria for
establishing the confidence intervals of the MOS (BT.500, P.910 and
P.913). The standardized procedures are not without their own limitations.
For example, in BT.500, if a subject is deemed an outlier, all the
raw opinion scores from that subject are discarded, which could be
an overkill. The BT.500 procedure also incorporates a number of hard
coded thresholds, which may not be suited for all test conditions.

As an alternative, we propose a simple model to account for two of
the most dominant behaviors of test subject inaccuracy: \emph{bias}
and \emph{inconsistency}. In addition, this model can effectively
deal with inattentive subject outliers that give random scores. Compared
to the BT.500-style subject rejection, the proposed model can be considered
as performing \emph{``soft'' subject rejection}, as it explicitly
models the subject outliers as having large inconsistencies, and thereby
limiting their contributions to the estimated quality score through
consistency weighting. To solve for the model parameters, we propose
to jointly optimize the likelihood function, also known as maximum
likelihood estimation (MLE) \cite{mle}. We present two numeric solvers:
1) a Newton-Raphson (NR) solver \cite{newtonraphson}, and 2) an
Alternating Projection (AP) solver. Compared to the NR solver which
was originally developed in \cite{dcc_arxiv}, \emph{the AP solver
is faster and more intuitive}. We further show that the AP solver
generalizes the P.913 post-test screening procedure by weighing a
subject's contribution to the true quality score by her consistency
(thus, the quality scores estimated can be thought as \emph{bias-subtractted
consistency-weighted MOS}). The AP solver also has the advantage of
\emph{having no hard coded parameters and thresholds}.

One of the challenges is to fairly compare the proposed methods to
its alternatives. To this end, we evaluate the proposed simple model
and its numerical solvers separately. To evaluate the model's fit
to real datasets, we use Bayesian Information Criterion (BIC) \cite{bic},
where the winner can be characterized as \emph{having a good fit to
data while maintaining a small number of parameters}. We also compare
the confidence intervals of the estimated quality scores, where a
tighter confidence interval implies a higher confidence in the estimation.
To evaluate the model's robustness against subject outliers, we perform
a simulation study on how the true quality score's root mean squared
error (RMSE) changes compared to the clean case as the number of outliers
increases. To validate that the numerical solvers are indeed accurate,
we use synthetic data to compare the recovered parameters against
the ground truth. Lastly, we show that the proposed methods are the
most valuable when the test conditions are challenging, by showing
their advantages in a crowdsourcing test and a cross-lab study.

The rest of the paper is organized as follows. In Section \ref{sec:Prior-Art-and}
we discuss the prior art and standards. We present the proposed model
in Section \ref{sec:Proposed-Model}, and then describe the two numerical
solvers in Section \ref{sec:Proposed-Solvers}. In Section \ref{sec:Confidence-Interval},
we discuss two alternative ways to calculate the confidence intervals
and compare their pros and cons. In Section \ref{sec:Experimental-Results}
we present the experimental results.

The code of this work is open-sourced on Github \cite{sureal}.

\section{Prior Art and Standards\label{sec:Prior-Art-and}}

Raw opinion scores collected from subjective experiments are known
to be influenced by the inherently noisy and unreliable nature of
human test subjects \cite{Hossfeld2011}. To compensate for the influence
of individuals, a common practice is to average the raw opinion scores
from multiple subjects, yielding a MOS per stimulus. Standardized
recommendations incorporate more advanced corrective mechanisms to
further compensate for test subjects\textquoteright{} influence, and
criteria for establishing the confidence intervals of MOS.
\begin{itemize}
\item ITU-R BT.500 Recommendation \cite{bt500} defines methodologies such
as single-stimulus continuous quality evaluation (SSCQE), double-stimulus
impairment scale (DSIS) and double-stimulus continuous quality scale
(DSCQS), and a corresponding procedure for subject rejection (ITU-R
BT.500-14 Section A1-2.3.1) prior to the calculation of MOS. Video
by video, the procedure counts the number of instances when a subject\textquoteright s
opinion score deviates by a few sigmas (i.e. standard deviation),
and rejects the subject if the occurrences are more than a fraction.
All scores corresponding to the rejected subjects are discarded, which
could be considered an overkill. On the other hand, our experiment
shows that, in the presence of many outlier subjects, the procedure
is only able to identify a fraction of them. In Section \ref{subsec:Visual-Examples},
we explain why this happens using a real example. Another of its drawbacks
is that the incorporation of a number of hard coded parameters and
thresholds to determine the outliers, which may not be suitable for
all conditions. The recommendation also establishes the corresponding
way of calculating the confidence interval (ITU-R BT.500-14 Section
A1-2.2.1).
\item ITU-T P.910 Recommendation \cite{p910} defines methodologies including
absolute category rating (ACR), degradation category rating DCR (equivalent
to DSIS), absolute category rating with hidden reference (ACR-HR)
and the corresponding differential MOS (DMOS) calculation, and recommends
using the BT.500 subject rejection and confidence interval calculation
procedure in conjunction.
\item ITU-T P.913 Recommendation \cite{p913} defines a procedure to remove
subject bias (ITU-T P.913 Section 12.4) before carrying out other
steps. It first finds the mean score per stimulus, and subtracts it
from the raw opinion scores to get the residual scores. Then it averages
the residue scores on a per-subject basis to yield an estimate of
each subject\textquoteright s bias. The biases are then removed from
the raw opinion scores. For P.913 to possess resistance to subject
outliers, it needs to be combined with a subject rejection strategy.
P.913 recommends several ways to do so but does not mandate one (ITU-T
P.913 03/2016 Section 11.4). For simplicity and consistency, in this
work, we use the same one as BT.500. Yet, by doing so, it inherits
similar weaknesses aforementioned.
\end{itemize}
For completeness, in Appendix \ref{sec:Appendix:-Mathematical-Descripti},
we give mathematical descriptions of the subject rejection method
standardized in ITU-R BT.500-14 and the subject bias removal method
in ITU-T P.913. 

\section{Proposed Model\label{sec:Proposed-Model}}

We propose a simple yet effective model to account for two of the
most dominant effects of test subject inaccuracy: subject bias and
subject inconsistency. The model is a simplified version of \cite{dcc_arxiv}
without considering the ambiguity of video content. Compared to the
previously proposed model, the solutions to the simplified model are
more efficient and stable.

Let $u_{ijr}$ be the opinion score voted by subject $i$ on stimulus
$j$ in repetition $r$. We assume that each opinion score $u_{ijr}$
can be represented by a random variable as follows:

\begin{equation}
U_{ijr}=\psi_{j}+\Delta_{i}+\upsilon_{i}X,\label{eq:model}
\end{equation}
where $\psi_{j}$ is the true quality of stimulus $j$, $\Delta_{i}$
represents the bias of subject $i$, the non-negative term $\upsilon_{i}$
represents the inconsistency of subject $i$, and $X\sim N(0,1)$
are i.i.d. Gaussian random variables. The index $r$ represents repetitions.

It is important to point out that a subject with erroneous behaviors
can be modeled by a large inconsistency value $\upsilon_{i}$. The
erroneous behaviors that can be modeled include but are not limited
to: subject giving random scores, subject being absent-minded for
a fraction of a session, or software issue that randomly shuffles
a subject\textquoteright s scores among multiple stimuli. By successfully
estimating $\upsilon_{i}$ and accounting its effect to calculating
the true quality score, we can compensate for subject outliers without
invoking BT.500-style subject rejection.

Given a collection of opinion scores $\{u_{ijr}\}$ from a subjective
experiment, the task is to solve for the free parameters $\theta=(\{\psi_{j}\},\{\Delta_{i}\},\{\upsilon_{i}\})$,
such that the model fits the observed scores the best. This can be
formulated as a maximum likelihood estimation (MLE) problem. Let the
log-likelihood function be 
\[
L(\theta)=\log P(\{u_{ijr}\}|\{\psi_{j}\},\{\Delta_{i}\},\{\upsilon_{i}\}),
\]
i.e. a monotonic measure of the probability of observing the given
raw scores, for a set of these parameters. We can solve the model
by finding $\hat{\theta}$ that maximizes $L(\theta)$, or $\hat{\theta}=\arg\max L(\theta)$.
This problem can be numerically solved by the Newton-Raphson method
or the Alternating Projection method, to be discussed in Section \ref{sec:Proposed-Solvers}.

It is important to notice that the recoverability of $\{\psi_{j}\}$
and $\{\Delta_{i}\}$ in (\ref{eq:model}) is up to a constant shift.
Formally, assume $\hat{\theta}=(\{\hat{\psi}_{j}\},\{\hat{\Delta}_{i}\},\{\hat{\upsilon}_{i}\})$
is a solution that maximizes $L(\theta)$, one can easily show that
$(\{\hat{\psi}_{j}+C\},\{\hat{\Delta}_{i}-C\},\{\hat{\upsilon}_{i}\})$
where $C\in\mathbb{R}$, is another solution that achieves the same
maximum likelihood value $L(\hat{\theta})$. This implies that the
optimal solution is not unique. In practice, we can enforce a unique
solution, by adding a constraint that forces the mean subject bias
to be zero, or 
\[
\sum_{i}\Delta_{i}=0.
\]
This intuitively makes sense, since bias is relative - saying everyone
is positively biased is equivalent to saying that no one is positively
biased. It is also equivalent to assuming that the sample of observers
that offer opinion scores in a subjective experiment are truly random
and do not consist of only ``expert'' viewers or ``lazy'' viewers
that tend to offer lower or higher opinion scores, as a whole. There
is always the possibility, once a subjective test establishes that
the population from where subjects were recruited have such a collective
bias, to change the condition and thus properly estimate what the
true ``typical'' observer, drawn from a more representative pool
that would vote.

Lastly, one should keep in mind that it is always possible to use
more complicated models than (\ref{eq:model}) to capture other effects
in a subjective experiment. For example, \cite{dcc_arxiv} considers
content ambiguity, and \cite{Janowski2014,janowski15} considers
per-stimulus ambiguity. There are also environment-related factors
that could induce biases. Additionally, the votes are influenced by
the voting scales chosen, for example, continuous vs. discrete \cite{janowski2019generalized}.
Our hope is that the proposed model strikes a good balance between
the model complexity and explanatory power. In Section \ref{subsec:Validation-using-Bayesian},
we show that the proposed model yields better model-data fit than
the BT.500 and P.913 being used today.

\section{Proposed Solvers\label{sec:Proposed-Solvers}}

Let us start by simplifying the form of the log-likelihood function
$L(\theta)$. We can write:

\begin{eqnarray}
L(\theta) & = & \log P(\{u_{ijr}\}|\{\psi_{j}\},\{\Delta_{i}\},\{\upsilon_{i}\})\nonumber \\
 & = & \log\prod_{ijr}P(u_{ijr}|\psi_{j},\Delta_{i},\upsilon_{i})\label{eq:independence-assumption}\\
 & = & \sum_{ijr}\log P(u_{ijr}|\psi_{j},\Delta_{i},\upsilon_{i})\nonumber \\
 & = & \sum_{ijr}\log f(u_{ijr}|\psi_{j}+\Delta_{i},\upsilon_{i})\nonumber \\
 & \cong & \sum_{ijr}-\log\upsilon_{i}-\frac{(u_{ijr}-\psi_{j}-\Delta_{i})^{2}}{2\upsilon_{i}^{2}}\label{eq:simplified-gaussian}
\end{eqnarray}
where (\ref{eq:independence-assumption}) uses the independence assumption
on opinion scores, $f(x|\mu,\upsilon)$ denotes the Gaussian density
function with mean $\mu$ and standard deviation $\upsilon$, and
$\cong$ denotes omission of constant terms.

Note that not every subject needs to vote on each stimulus in every
repetition. Our proposed solvers can effectively deal with subjective
tests with incomplete data where some observations $u_{ijr}$ are
missing. Denote by $\star$ the missing observations in an experiment.
All summations in this paper are ignoring the missing observations
$\star$, that is, $\sum_{ijr}$ is equivalent to $\sum_{ijr:u_{ijr}\neq\star}$,
and so on.

\subsection{Newton-Raphson (NR) Solver}

With (\ref{eq:simplified-gaussian}), the first- and second-order
partial derivatives of $L(\theta)$ can be derived (see Section \ref{sec:Appendix:partial-derivatives}).
We can apply the Newton-Raphson rule \cite{newtonraphson} $a^{new}\leftarrow a-\frac{\partial L/\partial a}{\partial^{2}L/\partial a^{2}}$
to update each parameter $a$ in iterations. We further use a refresh
rate parameter $\alpha$ to control the innovation rate to avoid overshooting.
Note that other update rules can be applied, but using the Newton-Raphson
rule yields nice interpretability.

Also note that the NR solver finds a local optimal solution when the
problem is non-convex. It is important to initialize the parameters
properly. We choose zeros as the initial values for $\{\Delta_{i}\}$,
the mean score $MOS_{j}=(\sum_{ir}1)^{-1}\sum_{ir}u_{ijr}$ for $\{\psi_{j}\}$,
and the residue standard deviation $RSD_{i}=\sigma_{i}(\{\epsilon_{ijr}\})$
for $\{\upsilon_{i}\}$, where $\epsilon_{ijr}=u_{ijr}-MOS_{j}$ is
the ``residue'', $\sigma_{i}(\{\epsilon_{ijr}\})=\sqrt{(\sum_{jr}1)^{-1}\sum_{jr}(\epsilon_{ijr}-\epsilon_{i})^{2}}$,
and $\epsilon_{i}=(\sum_{jr}1)^{-1}\sum_{jr}\epsilon_{ijr}$. The
NR solver is summarized in Algorithm \ref{nr_solver}. A good choice
of innovation rate and stop threshold are $\alpha=0.1$ and $\psi^{thr}=1e^{-9}$,
respectively, but varying these parameters would not significantly
change the result.

\begin{algorithm}[t]
\normalsize
\begin{itemize}
\item Input: 

\begin{itemize}
\item $u_{ijr}$ for subject $i=1,...,I$, stimulus $j=1,...,J$ and repetition
$r=1,...,R$.
\item Refresh rate $\alpha$.
\item Stop threshold $\psi^{thr}$.
\end{itemize}
\item Initialize $\{\Delta_{i}\}\leftarrow\{0\}$, $\{\psi_{j}\}\leftarrow\{MOS_{j}\}$,
$\{\upsilon_{i}\}\leftarrow\{RSD_{i}\}$. 
\item Loop:

\begin{itemize}
\item $\{\psi_{j}^{prev}\}\leftarrow\{\psi_{j}\}$.
\item $\Delta_{i}\leftarrow(1-\alpha)\cdot\Delta_{i}+\alpha\cdot\Delta_{i}^{new}$
where $\Delta_{i}^{new}=\Delta_{i}-\frac{\partial L(\theta)/\partial\Delta_{i}}{\partial^{2}L(\theta)/\partial\Delta_{i}^{2}}$
for $i=1,...,I$.
\item $\upsilon_{i}\leftarrow(1-\alpha)\cdot\upsilon_{i}+\alpha\cdot\upsilon_{i}^{new}$
where $\upsilon_{i}^{new}=\upsilon_{i}-\frac{\partial L(\theta)/\partial\upsilon_{i}}{\partial^{2}L(\theta)/\partial\upsilon_{i}^{2}}$
for $i=1,...,I$.
\item $\psi_{j}\leftarrow(1-\alpha)\cdot\psi_{j}+\alpha\cdot\psi_{j}^{new}$
where $\psi_{j}^{new}=\psi_{j}-\frac{\partial L(\theta)/\partial\psi_{j}}{\partial^{2}L(\theta)/\partial\psi_{j}^{2}}$
for $j=1,...,J$.
\item If $\left(\sum_{j=1}^{J}(\psi_{j}-\psi_{j}^{prev})^{2}\right)^{\frac{1}{2}}<\psi^{thr}$,
break.
\end{itemize}
\item Output: $\{\psi_{j}\}$, $\{\Delta_{i}\}$, $\{\upsilon_{i}\}$.
\end{itemize}
\caption{Proposed Newton-Raphson (NR) solver}
\label{nr_solver}
\end{algorithm}

The ``new'' parameters can be simplified to the following form:
\begin{eqnarray}
\psi_{j}^{new} & = & \frac{\sum_{ir}\upsilon_{i}^{-2}(u_{ijr}-\Delta_{i})}{\sum_{ir}\upsilon_{i}^{-2}},\label{eq:nr-psi}\\
\Delta_{i}^{new} & = & \frac{\sum_{jr}(u_{ijr}-\psi_{j})}{\sum_{jr}1},\label{eq:nr-delta}\\
\upsilon_{i}^{new} & = & \upsilon_{i}\frac{\sum_{jr}2\upsilon_{i}^{2}-4(u_{ijr}-\psi_{j}-\Delta_{i})^{2}}{\sum_{jr}\upsilon_{i}^{2}-3(u_{ijr}-\psi_{j}-\Delta_{i})^{2}}.\nonumber 
\end{eqnarray}
Note that there are strong intuitions behind the expressions for the
newly estimated true quality $\psi_{j}^{new}$ and subject bias $\Delta_{i}^{new}$.
In each iteration, $\psi_{j}^{new}$ is re-estimated, as the weighted
mean of the opinion scores $u_{ijr}$ with the currently estimated
subject bias $\Delta_{i}$ removed. Each opinion score is weighted
by the ``subject consistency'' $\upsilon_{i}^{-2}$, i.e., the higher
the inconsistency for subject $i$, the less reliable the opinion
score, hence less the weight. For the subject bias $\Delta_{i}^{new}$,
it is simply the average shift between subject $i$'s opinion scores
and the true values.

\subsection{Alternating Projection (AP) Solver}

This solver is called ``alternating projection'' because in a loop,
it alternates between projecting (or averaging) the opinion scores
along the subject dimension and the stimulus dimension. To start,
we initialize $\{\psi_{j}\}$ to $\{MOS_{j}\}$, where $MOS_{j}=(\sum_{ir}1)^{-1}\sum_{ir}u_{ijr}$,
same as the NR solver. The subject bias $\{\Delta_{i}\}$ is initialized
differently to $\{BIAS_{i}\}$, where $BIAS_{i}=(\sum_{jr}1)^{-1}\sum_{jr}(u_{ijr}-MOS_{j})$
is the average shift between subject $i$'s opinion scores and the
true values. Note that the calculation of $\{MOS_{j}\}$ and $\{BIAS_{i}\}$
matches precisely to the ones in Algorithm \ref{subjbiasrmv} (ITU-T
P.913). Within the loop, first, the ``residue'' $\text{\ensuremath{\epsilon_{ijr}}}$
is updated, followed by the calculation of the subject inconsistency
$\upsilon_{i}$ as the residue's standard deviation per subject $\sigma_{i}(\{\epsilon_{ijr}\})$,
with:
\begin{eqnarray}
\sigma_{i}(\{\epsilon_{ijr}\}) & = & \sqrt{(\sum_{jr}1)^{-1}\sum_{jr}(\epsilon_{ijr}-\epsilon_{i})^{2}},\label{eq:sigma_i}\\
\epsilon_{i} & = & (\sum_{jr}1)^{-1}\sum_{jr}\epsilon_{ijr}.\label{eq:epsilon_i}
\end{eqnarray}

Then, the true quality $\{\psi_{j}\}$ and the subject bias $\{\Delta_{i}\}$
are re-estimated, by averaging the opinion scores along either the
subject dimension $i$ or the stimulus dimension $j$. The projection
formula precisely matches equations (\ref{eq:nr-psi}) and (\ref{eq:nr-delta})
of the Newton-Raphson method. The AP solver is summarized in Algorithm
\ref{ap_solver}. A good choice of the stop threshold is $\psi^{thr}=1e^{-8}$.

The AP solver generalizes the P.913 post-test screening (Section 12.4)
in the following sense. First, the AP solver is iterative until convergence
whereas P.913 only goes through the initialization steps. Second,
in the AP solver, the re-estimation of quality score $\psi_{j}$ is
weighted by the subject consistency $\upsilon_{i}^{-2}$ whereas in
P.913, the re-estimation is unweighted. Please note that weighting
multiple random variables by the inverse of their variance is the
minimum error parameter estimation, as can be trivially proven through
Lagrange multipliers. Intuitively, the estimated quality scores of
the AP solver can be regarded as \emph{bias-subtracted consistency-weighted
MOS}.

\begin{algorithm}[t]
\normalsize
\begin{itemize}
\item Input: 

\begin{itemize}
\item $u_{ijr}$ for subject $i=1,...,I$, stimulus $j=1,...,J$ and repetition
$r=1,...,R$.
\item Stop threshold $\psi^{thr}$.
\end{itemize}
\item Initialize $\{\psi_{j}\}\leftarrow\{MOS_{j}\}$, $\{\Delta_{i}\}\leftarrow\{BIAS_{i}\}$.

\item Loop:

\begin{itemize}
\item $\{\psi_{j}^{prev}\}\leftarrow\{\psi_{j}\}$.
\item $\epsilon_{ijr}=u_{ijr}-\psi_{j}-\Delta_{i}$ for $i=1,...,I$, $j=1,...,J$
and $r=1,...,R$.
\item $\upsilon_{i}\leftarrow\sigma_{i}(\{\epsilon_{ijr}\})$ for $i=1,...,I$.
\item $\psi_{j}\leftarrow\frac{\sum_{ir}\upsilon_{i}^{-2}(u_{ijr}-\Delta_{i})}{\sum_{ir}\upsilon_{i}^{-2}}$
for $j=1,...,J$.
\item $\Delta_{i}\leftarrow\frac{\sum_{jr}(u_{ijr}-\psi_{j})}{\sum_{jr}1},$
for $i=1,...,I$.
\item If $\left(\sum_{j=1}^{J}(\psi_{j}-\psi_{j}^{prev})^{2}\right)^{\frac{1}{2}}<\psi^{thr}$,
break.
\end{itemize}
\item Output: $\{\psi_{j}\}$, $\{\Delta_{i}\}$, $\{\upsilon_{i}\}$.
\end{itemize}
\caption{Proposed Alternating Projection (AP) solver}
\label{ap_solver}
\end{algorithm}

\section{Confidence Interval\label{sec:Confidence-Interval}}

The estimation of each model parameter $\{\psi_{j}\}$, $\{\Delta_{i}\}$,
$\{\upsilon_{i}\}$ is associated with a confidence interval (CI).
One way to calculate the CIs is through their asymptotic analytical
forms. Using the Cramer-Rao bound \cite{cover2006elements}, the
asymptotic $95\%$ confidence intervals for the mean terms $\psi_{j}$
and $\Delta_{i}$ have the form $CI(a)=a\pm1.96\left(-\frac{\partial^{2}L(a)}{\partial a^{2}}\right)^{-\frac{1}{2}}$,
where their second-order derivatives $\frac{\partial^{2}L(a)}{\partial a^{2}}$
can be found in Section \ref{sec:Appendix:partial-derivatives}. The
$95\%$ confidence interval for the standard deviation term $\upsilon_{i}$
has the form $\left(\sqrt{\frac{k}{\chi_{k}^{2}(0.975)}}\upsilon,\sqrt{\frac{k}{\chi_{k}^{2}(0.025)}}\upsilon\right)$,
where $\chi_{k}^{2}(a)$ is the percent point function (i.e. the inverse
of the cumulative distribution function) of a chi-square distribution
with $k$ degrees of freedom. After simplification, the confidence
intervals for $\psi_{j}$, $\Delta_{i}$ and $\upsilon_{i}$ are:

\begin{eqnarray}
CI(\psi_{j}) & = & \psi_{j}\pm1.96\frac{1}{\sqrt{\sum_{ir}\upsilon_{i}^{-2}}},\label{eq:quality_ci_mle}\\
CI(\Delta_{i}) & = & \Delta_{i}\pm1.96\frac{\upsilon_{i}}{\sqrt{\sum_{jr}1}},\nonumber \\
CI(\upsilon_{i}) & = & \left(\begin{array}{cc}
\sqrt{\frac{k_{i}}{\chi_{k_{i}}^{2}(0.975)}}\upsilon_{i}, & \sqrt{\frac{k_{i}}{\chi_{k_{i}}^{2}(0.025)}}\upsilon_{i}\end{array}\right),\nonumber 
\end{eqnarray}
where $k_{i}=\sum_{jr}1$ is the number of samples that subject $i$
has viewed.

There is one practical limitation with Equation (\ref{eq:quality_ci_mle}).
Recall that $\sum_{ir}$ is equivalent to $\sum_{ir:u_{ijr}\neq\star}$,
where $\star$ represents missing observation. If there is no missing
observations (i.e. every subject votes on every stimulus), then the
lengths of the confidence intervals for $\psi_{j}$, $j=1,...,J$
will be all equal to the same value $\frac{3.92}{\sqrt{\sum_{ir}\upsilon_{i}^{-2}}}$
(since it is independent of the subscript $j$). In other words, we
have equal confidence in the estimation of quality scores $\psi_{j}$,
$j=1,...,J$. Although this is theoretically correct (since all quality
scores are estimated jointly) and it results in the tightest CIs,
it deviates from a conventional approach (for example, plain MOS,
or BT.500), where each quality score has a different CI length (see
Section \ref{sec:Appendix:-An-MLE-MOS} for a MLE interpretation of
the plain MOS). Practically, it raises the concern that the CIs are
unable to capture the behaviors of individual stimuli. 

To address this concern, we propose an alternative calculation of
CI for $\psi_{j}$, which yields looser but discriminative CIs. Let
$\upsilon_{j}=\sigma_{j}(\{\epsilon_{ijr}\})$ denote the per-stimulus
variability, where $\sigma_{j}(\{\epsilon_{ijr}\})$ is the per-stimulus
standard deviation of the residues $\{\epsilon_{ijr}\}$, with:
\begin{eqnarray}
\sigma_{j}(\{\epsilon_{ijr}\}) & = & \sqrt{(\sum_{ir}1)^{-1}\sum_{ir}(\epsilon_{ijr}-\epsilon_{j})^{2}},\label{eq:sigma_j}\\
\epsilon_{j} & = & (\sum_{ir}1)^{-1}\sum_{ir}\epsilon_{ijr}.\label{eq:epsilon_j}
\end{eqnarray}
The alternative CI for $\psi_{j}$ (denoted by $CI_{2}$) can then
be calculated as:
\begin{equation}
CI_{2}(\psi_{j})=\psi_{j}\pm1.96\frac{\upsilon_{j}}{\sqrt{\sum_{ir}1}}.\label{eq:quality_ci_mle_alt}
\end{equation}
Comparing (\ref{eq:sigma_j}) (\ref{eq:epsilon_j}) with (\ref{eq:sigma_i})
(\ref{eq:epsilon_i}), one can see that the main difference is the
``projection direction'' in the tensor $\{\epsilon_{ijr}\}$ when
calculating the mean and the standard deviation. Through simulations,
we will show in Section \ref{subsec:Validation-of-Solvers} that $CI_{2}$
is also theoretically correct, although it yields less tight CIs.
In the following sections, the AP solver combined with $CI_{2}$ is
denoted by AP2.

\begin{figure}
\begin{centering}
\includegraphics[scale=0.3]{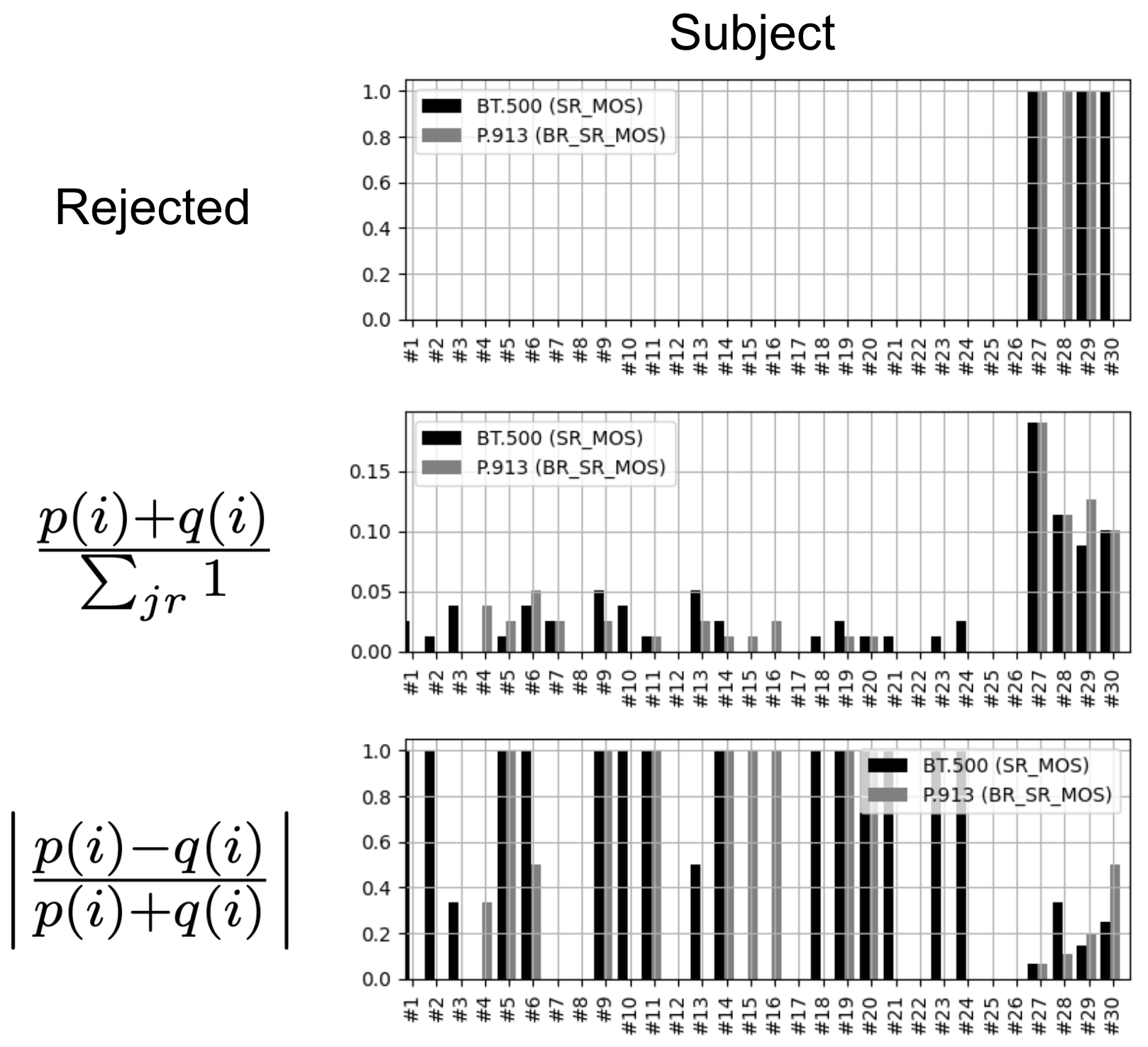}
\par\end{centering}
\caption{\label{fig:bt500_913_sr} Subject rejection result by BT.500 and P.913
on the NFLX Public dataset. The top plot shows the final rejection
result, where value 1.0 means rejected and 0.0 means not rejected.
The middle and bottom plots show the intermediate result of $(\sum_{jr}1)^{-1}(p(i)+q(i))$
and $\left|(p(i)+q(i))^{-1}(p(i)-q(i))\right|$, respectively. (SR:
subject rejection; BR: bias removal.)}
\end{figure}

\begin{figure*}[p]
\begin{centering}
\noindent\begin{minipage}[t]{0.9\columnwidth}%
\begin{center}
\includegraphics[width=1\columnwidth]{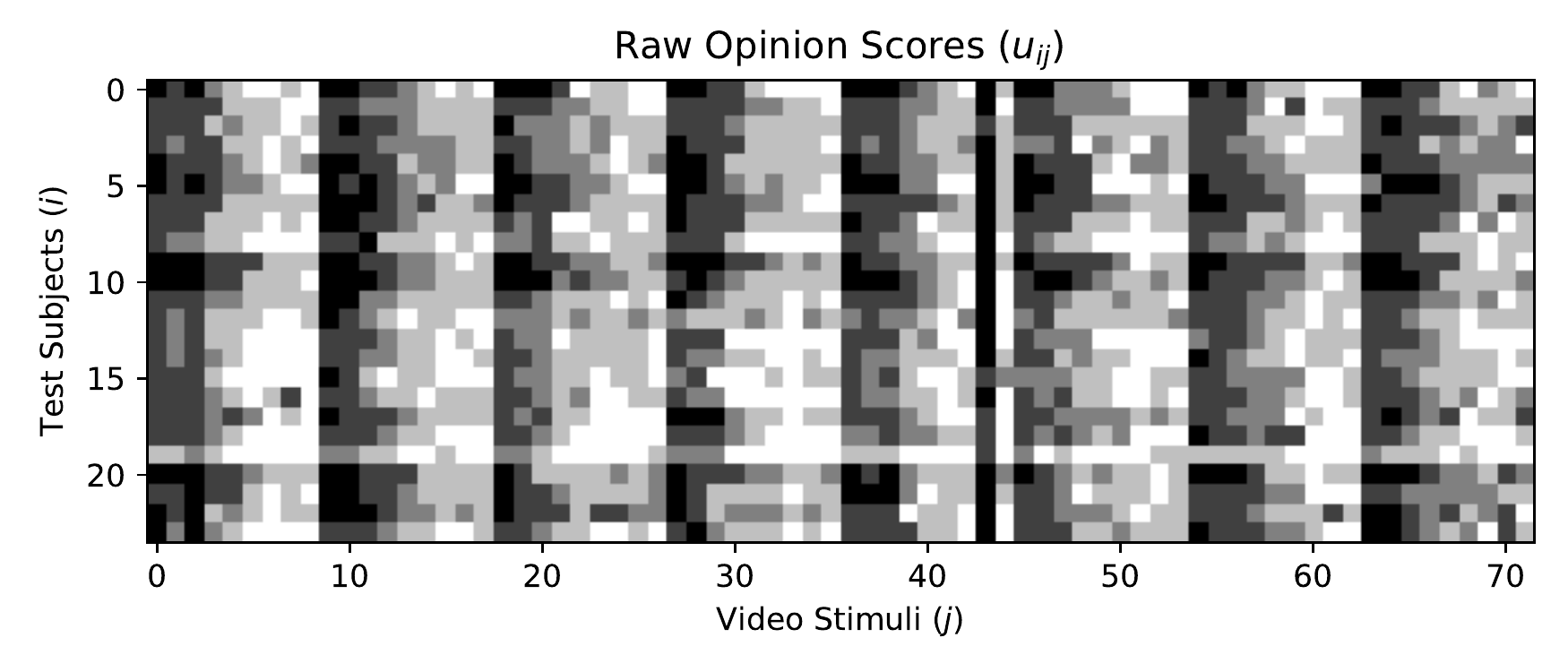}
\par\end{center}
\vspace{-0.2in}
\begin{center}
(a) VQEG HD3 dataset
\par\end{center}%
\end{minipage}
\noindent\begin{minipage}[t]{0.9\columnwidth}%
\begin{center}
\includegraphics[width=1\columnwidth]{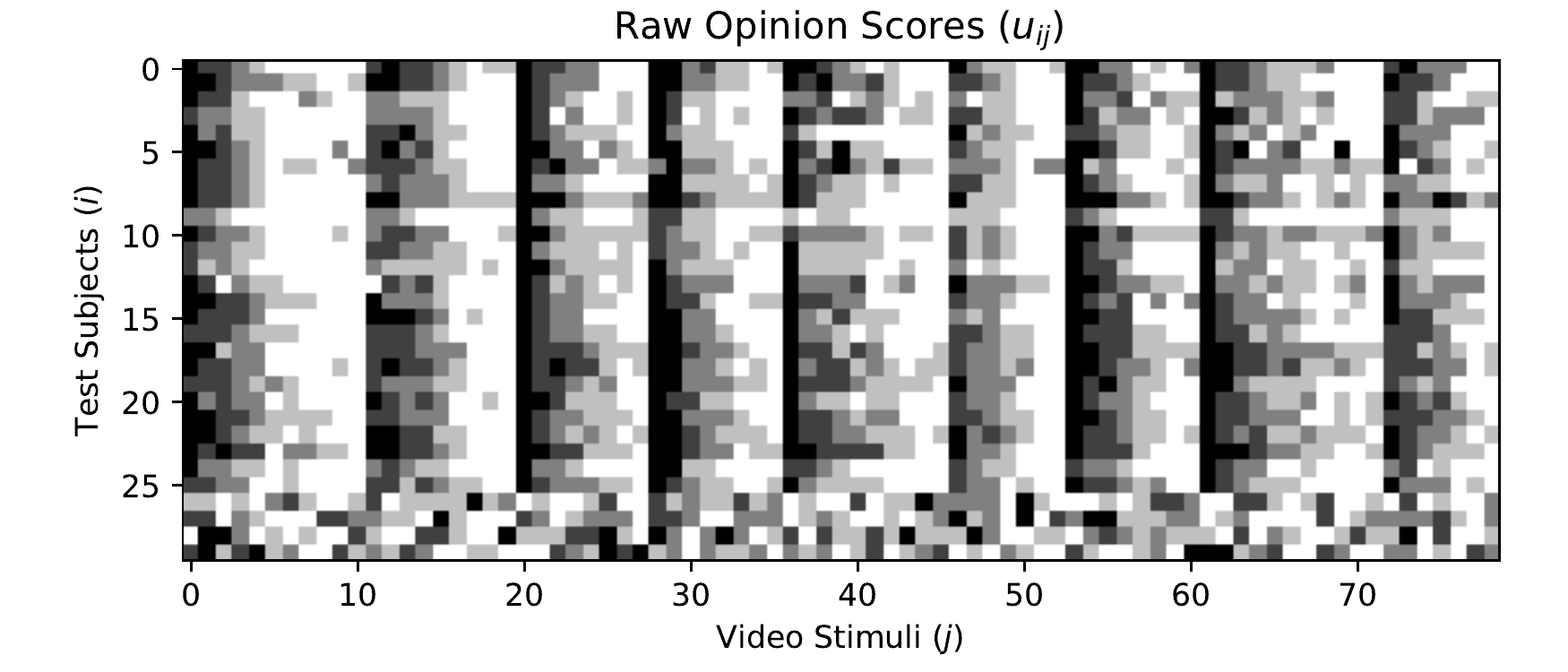}
\par\end{center}
\vspace{-0.2in}
\begin{center}
(b) NFLX Public dataset
\par\end{center}%
\end{minipage}
\caption{\label{fig:Raw-opinion-scores}Raw opinion scores from (a) the VQEG
HD3 dataset \cite{HDTV_Phase_I_test} and (b) the NFLX Public dataset
\cite{nflx-public}. Each pixel represents a raw opinion score. The
darker the color, the lower the score. The impaired videos are arranged
by contents, and within each content, from low quality to high quality
(with the reference video always appears last). For the NFLX Public
dataset, the last four rows correspond to corrupted subjective data.}
\par\end{centering}
\end{figure*}

\begin{figure*}[p]
\begin{centering}
\noindent\begin{minipage}[t]{1.3\columnwidth}%
\begin{center}
\includegraphics[width=1\columnwidth]{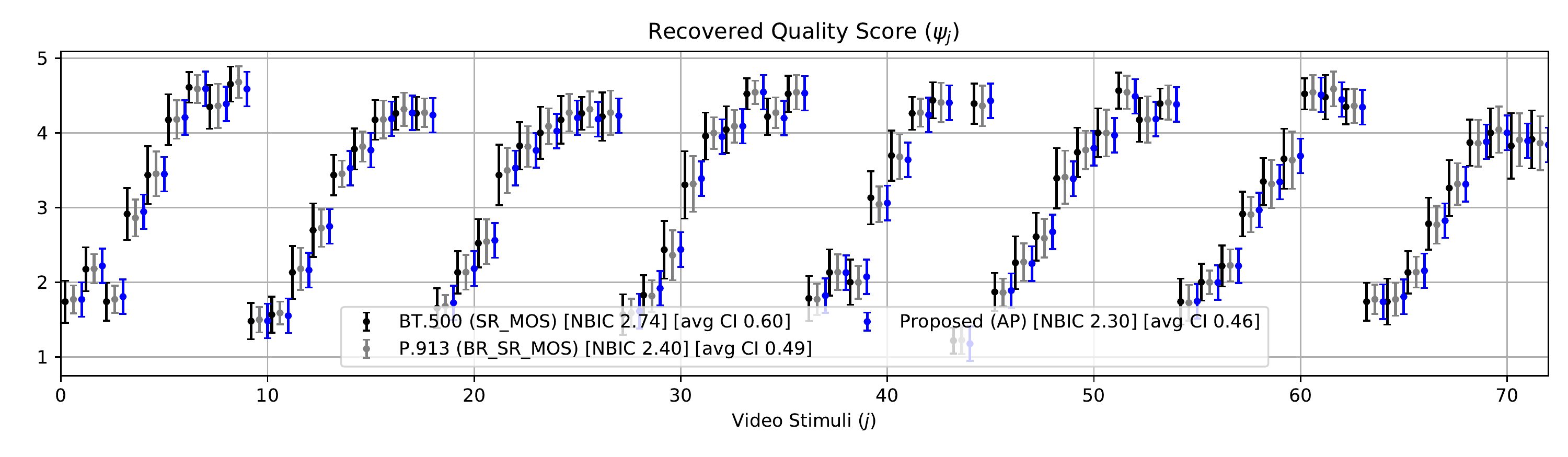}
\par\end{center}
\vspace{-0.2in}
\begin{center}
(a) VQEG HD3 dataset
\par\end{center}%
\end{minipage}
\par\end{centering}
\begin{centering}
\noindent\begin{minipage}[t]{1.3\columnwidth}%
\begin{center}
\includegraphics[width=1\columnwidth]{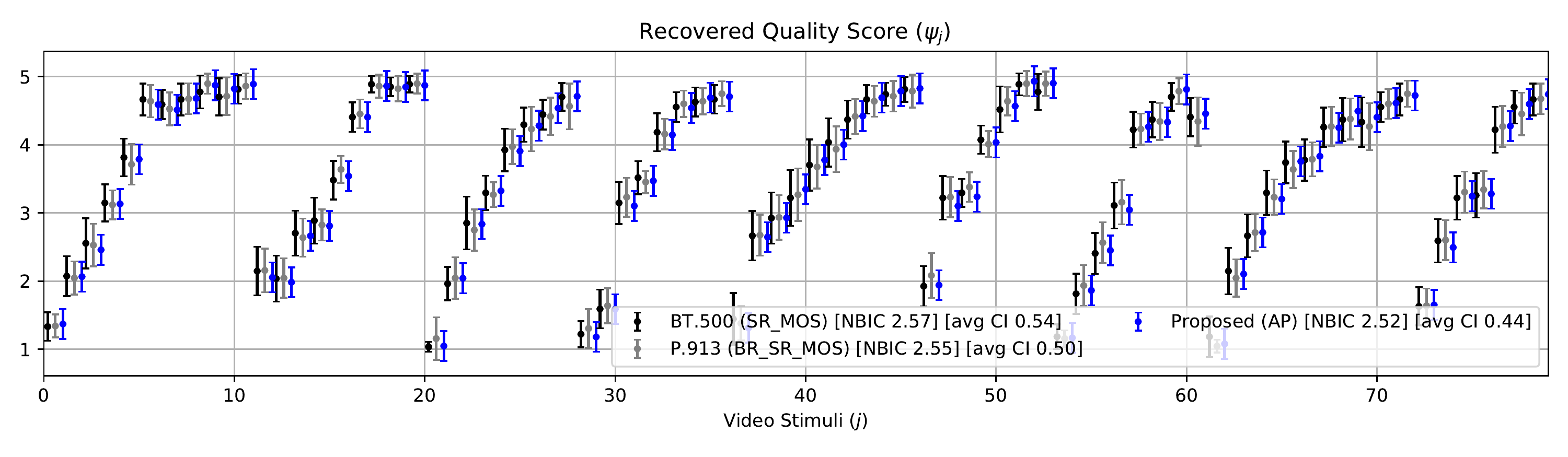}
\par\end{center}
\vspace{-0.2in}
\begin{center}
(b) NFLX Public dataset
\par\end{center}%
\end{minipage}
\par\end{centering}
\caption{\label{fig:quality-scores}Recovered quality score $\psi_{j}$ and
its confidence interval (based on (\ref{eq:quality_ci_mle})) for
the four methods compared, on (a) the VQEG HD3 dataset and (b) the
NFLX Public dataset. The proposed NR method is not shown in the plots
since it virtually produces identical results as the proposed AP method.
For each method compared, the NBIC score (see Section \ref{subsec:Validation-using-Bayesian})
and the average length of the confidence interval are reported. (SR:
subject rejection; BR: bias removal; avg CI: average confidence interval;
NBIC: Bayesian Information Criterion; NR: Newton-Raphson; AP: Alternating
Projection.)}
\end{figure*}

\begin{figure*}[p]
\begin{centering}
\begin{minipage}[t]{1\columnwidth}%
\begin{center}
\includegraphics[width=0.9\columnwidth]{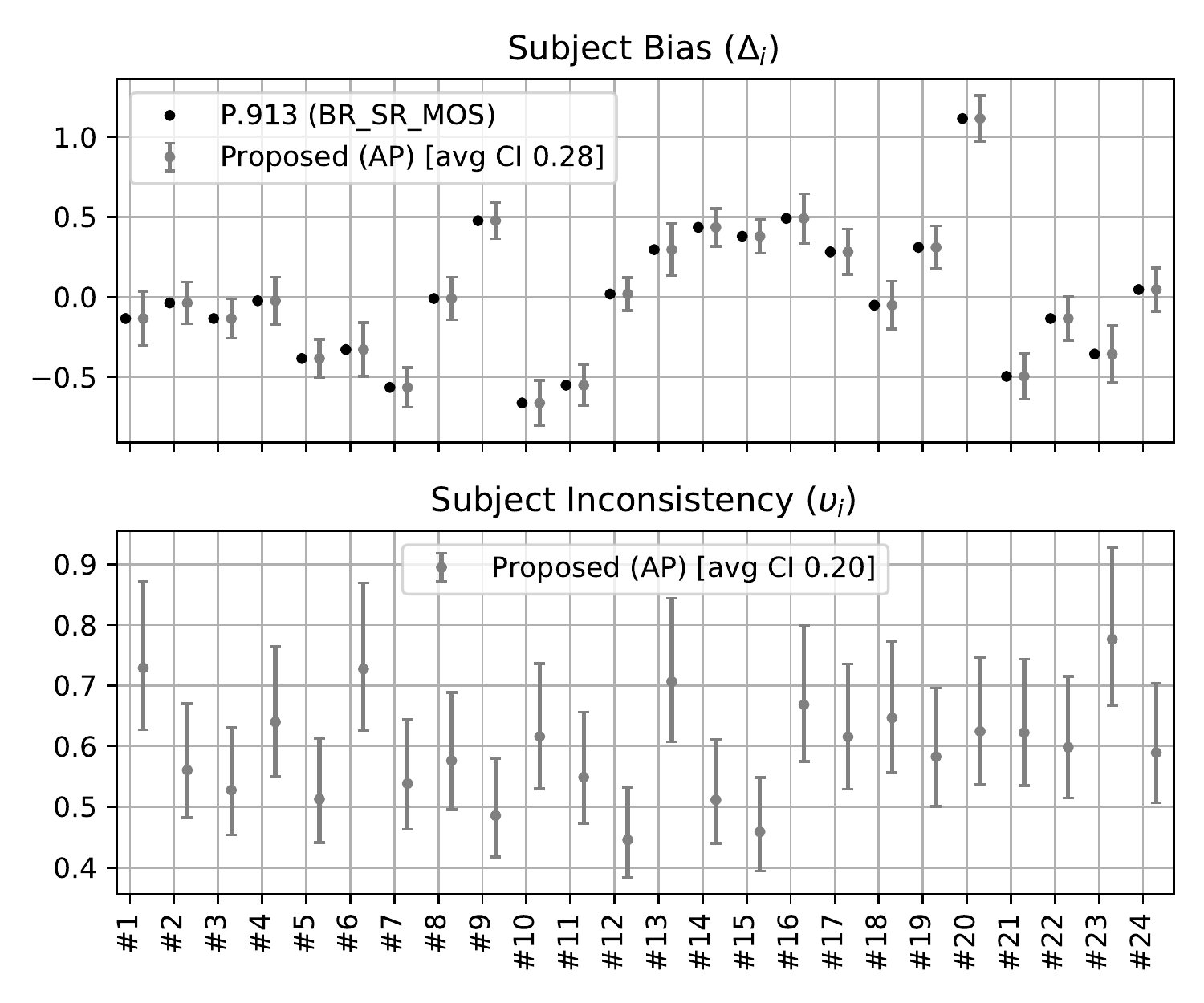}
\par\end{center}
\vspace{-0.2in}
\begin{center}
(a) VQEG HD3 dataset
\par\end{center}%
\end{minipage}%
\begin{minipage}[t]{1\columnwidth}%
\begin{center}
\includegraphics[width=0.9\columnwidth]{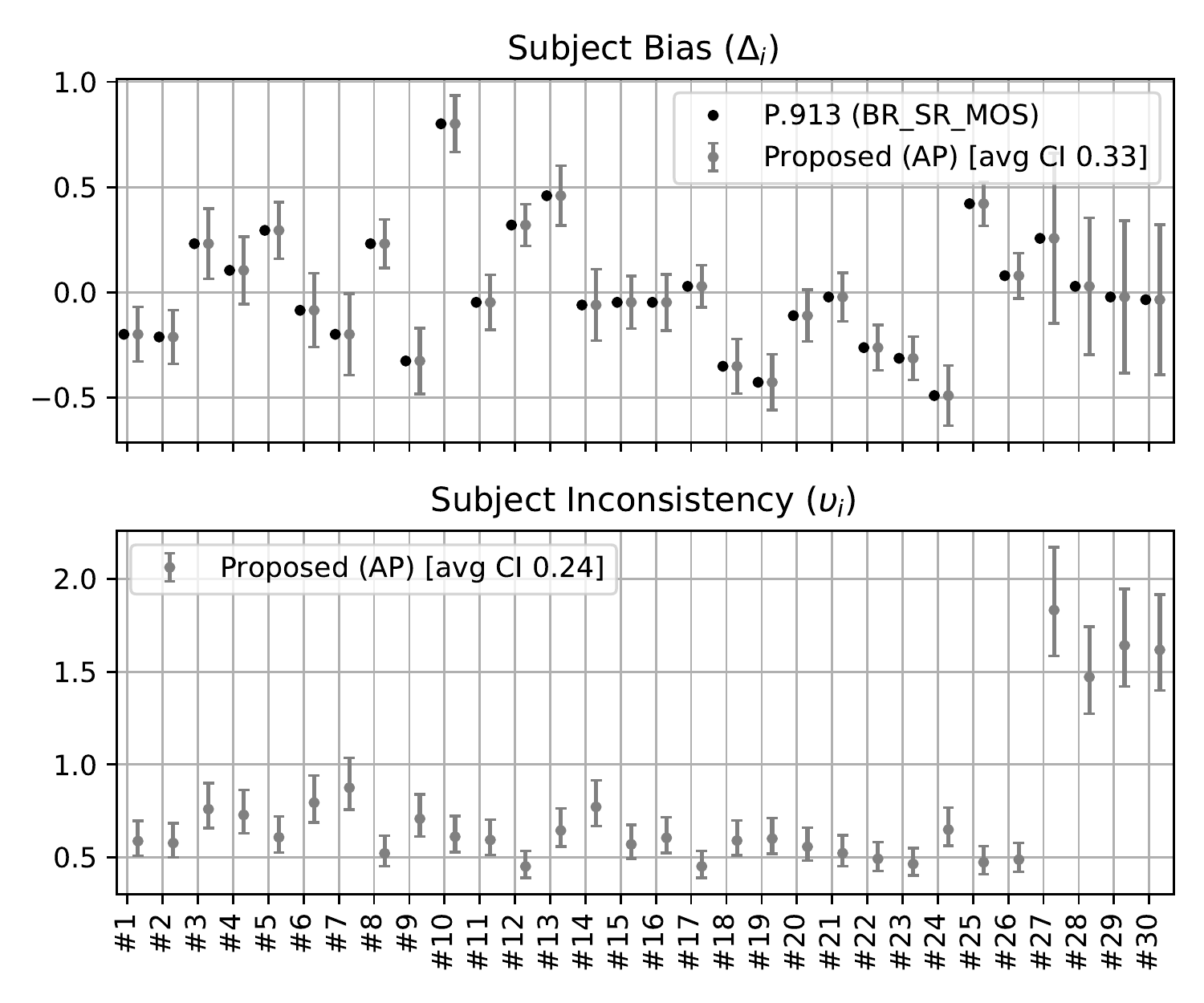}
\par\end{center}
\vspace{-0.2in}
\begin{center}
(b) NFLX Public dataset
\par\end{center}%
\end{minipage}
\par\end{centering}
\caption{\label{fig:subject-scores}Recovered subject bias $\Delta_{i}$ and
subject inconsistency $\upsilon_{i}$ for each subject $i$, for the
methods compared, on (a) the VQEG HD3 dataset and (b) the NFLX Public
dataset. The proposed NR method is not shown in the plots since it
virtually produces identical results as the proposed AP method. For
each method compared, the average length of the confidence interval
is reported. (SR: subject rejection; BR: bias removal; avg CI: average
confidence interval; NR: Newton-Raphson; AP: Alternating Projection.)}
\end{figure*}

\section{Experimental Results\label{sec:Experimental-Results}}

We compare the proposed method (the proposed model and its two numerical
solvers) with the prior art BT.500 and P.913 recommendations. We first
illustrate the proposed model by giving visual examples on two datasets:
VQEG HD3 dataset \cite{HDTV_Phase_I_test} (which is the compression-only
subset of the larger HDTV Ph1 Exp3 dataset) and the NFLX Public dataset
\cite{nflx-public}. We then validate the model-data fit using the
Bayesian Information Criterion (BIC) on 22 datasets, including 20
datasets as part of a different larger experiments: VQEG HDTV Phase
I \cite{HDTV_Phase_I_test}; ITS4S \cite{ITS4S_test}; AGH/NTIA
\cite{Janowski2014,AGH_NTIA_14-505}; MM2 \cite{MM2_test}; ITU-T
Supp23 Exp1 \cite{ITU-TSupp23}; and ITS4S2 \cite{ITS4S2_test}.
We also evaluate the confidence intervals on the estimated quality
scores on these 22 datasets. Next, we demonstrate that the proposed
model is much more effective in dealing with outlier subjects. Lastly,
we use synthetic data to validate the accuracy of the numerical solvers
and the confidence interval calculation. Lastly, we show that the
proposed method is most valuable when the test conditions are challenging,
by showing their advantages in a crowdsourcing test conducted at Netflix
and the VQEG FRTV Phase I cross-lab study \cite{frtvp1}.

\subsection{Visual Examples\label{subsec:Visual-Examples}}

First, we demonstrate the proposed method on the VQEG HD3 and the
NFLX Public datasets. Refer to Figure \ref{fig:Raw-opinion-scores}
for a visualization of the raw opinion scores. The 44th video of the
VQEG HD3 dataset has a quality issue such that all its scores are
low. The NFLX Public dataset includes four subjects whose raw scores
were shuffled due to a software issue during data collection.

Figure \ref{fig:quality-scores} shows the recovered quality scores
of the four methods compared. The quality scores recovered by the
two proposed methods are numerically different from the ones of BT.500
and P.913, suggesting that the recovery is non-trivial. The average
confidence intervals (based on (\ref{eq:quality_ci_mle})) by the
proposed methods are generally tighter, compared to the ones of BT.500
and P.913, suggesting that the estimation has higher confidence. Due
to clutteredness, we do not plot the alternative $CI_{2}$ (\ref{eq:quality_ci_mle_alt})
in Figure \ref{fig:quality-scores}, but will show them in Section
\ref{subsec:Confidence-Interval-of}. The NBIC scores, to be discussed
in detail in Section \ref{subsec:Validation-using-Bayesian}, represent
how well the model fits the data. It can be observed that the proposed
model fits the data better than BT.500 and P.913.

Figure \ref{fig:subject-scores} shows the recovered subject bias
and subject inconsistency by the methods compared. On the VQEG HD3
dataset, it can be seen that the 20th subject has the most positive
bias, which is evidenced by the whitish horizontal strip visible in
Figure \ref{fig:Raw-opinion-scores} (a). On the NFLX Public dataset,
the last four subjects, whose raw scores are scrambled, have very
high subject inconsistency values. Correspondingly, their estimated
biases have very loose confidence intervals. This illustrates that
the proposed model is effective in modeling outlier subjects. 

The subject bias and inconsistency revealed through the recovery process
could be valuable information for subject screening. Unlike BT.500,
which makes a binary decision on if a subject is accepted/rejected,
the proposed approach characterizes a subject's inaccuracy in two
dimensions, along with their confidence intervals, allowing further
interpretation and study. How to use the bias and inconsistency information
to better screen subjects remains our future work.

\subsubsection{Comparison with BT.500 and P.913}

As a comparison, we also visualize the subject rejection results from
BT.500 and P.913 on the NFLX Public dataset, as shown in Figure \ref{fig:bt500_913_sr}.
These recommendations encode the hard rule that if 
\begin{equation}
\frac{p(i)+q(i)}{\sum_{jr}1}\geq0.05\label{eq:bt500_criterion_1}
\end{equation}
 and 
\begin{equation}
\left|\frac{p(i)-q(i)}{p(i)+q(i)}\right|<0.3,\label{eq:bt500_criterion_2}
\end{equation}
subject $i$ is rejected. Intuitively, (\ref{eq:bt500_criterion_1})
looks at the fraction of scores that are considered outliers. As shown
in the middle plot of Figure \ref{fig:bt500_913_sr}, for both BT.500
and P.913, all four outliers meet this criterion. The real problem
lies in (\ref{eq:bt500_criterion_2}), which says that subject $i$
is rejected \emph{only if} the distribution is not too skewed. The
rule itself seems benign, as it tests the skewness of the distribution
and only when the distribution is not too skewed, (\ref{eq:bt500_criterion_1})
should apply. But by this rule, for either BT.500 and P.913, only
three out of four outliers meet the hard-coded threshold of 0.3. This
example exposes the limitation of the hard-coded rules in BT.500 and
P.913. On the contrary, the proposed method does not use hard-coded
rules, thus is immune from this problem.

\subsection{Model-Data Fit\label{subsec:Validation-using-Bayesian}}

Bayesian Information Criterion (BIC) \cite{bic} is a criterion for
model-data fit. When fitting models, it is possible to increase the
likelihood by adding parameters, but doing so may result in overfitting.
BIC attempts to balance between the degree of freedom (characterized
by the number of free parameters) and the goodness of fit (characterized
by the log-likelihood function). Formally, BIC is defined as $BIC=\log(n)|\theta|-2L(\theta)$,
where $n=|\{u_{ijr}\}|$ is the total number of observations (i.e.
the number of opinion scores), $|\theta|$ is the number of model
parameters, and $L(\theta)$ is the log-likelihood function. One can
interpret that the lower the free parameter numbers $|\theta|$, and
the higher the log-likelihood $L(\theta)$, the lower the BIC, and
hence the better fit. In this work, we adopt the notion of a \emph{normalized
BIC} (NBIC), defined as the BIC divided by the number of observations,
or:
\[
NBIC=\frac{\log(n)|\theta|-2L(\theta)}{n},
\]
as the model fit criterion, for easier comparison across datasets
(since different datasets have a different $n$).

Table \ref{table:bic} shows the NBIC reported on the compared methods
on the 22 public datasets. The MOS method is the plain MOS without
subject rejection or subject bias removal. $|\theta|$ for MOS and
BT.500 is $2J$, where $J$ is the number of stimuli (refer to Section
\ref{sec:Appendix:-An-MLE-MOS} for a MLE interpretation of the plain
MOS). For P.913, $|\theta|$ is equal to $2J+I$, where $I$ is the
number of subjects (due to the subject bias term). For the calculation
of the log-likelihood function, notice that if subject rejection is
applied, only the opinion scores after rejection are taken into account.
The result in Table \ref{table:bic} shows that the proposed two solvers
yield better model-data fit than the plain MOS, BT.500 and P.913 approaches.

\begin{table}
\begin{centering}
\begin{tabular}{|c|cccc|}
\hline 
Dataset & MOS & BT.500 & P.913 & NR/AP\tabularnewline
\hline 
VQEG HD3 & 2.75 & 2.74 & 2.39 & \textbf{2.30}\tabularnewline
\hline 
NFLX Public & 2.97 & 2.57 & 2.55 & \textbf{2.52}\tabularnewline
\hline 
HDTV Ph1 Exp1 & 2.45 & 2.46 & 2.38 & \textbf{2.20}\tabularnewline
\hline 
HDTV Ph1 Exp2 & 2.72 & 2.72 & 2.52 & \textbf{2.32}\tabularnewline
\hline 
HDTV Ph1 Exp3 & 2.72 & 2.71 & 2.37 & \textbf{2.29}\tabularnewline
\hline 
HDTV Ph1 Exp4 & 2.96 & 2.96 & 2.51 & \textbf{2.27}\tabularnewline
\hline 
HDTV Ph1 Exp5 & 2.77 & 2.77 & 2.47 & \textbf{2.33}\tabularnewline
\hline 
HDTV Ph1 Exp6 & 2.51 & 2.49 & 2.32 & \textbf{2.16}\tabularnewline
\hline 
ITU-T Supp23 Exp1 & 2.91 & 2.91 & 2.35 & \textbf{2.31}\tabularnewline
\hline 
MM2 1 & 2.80 & 2.78 & 2.83 & \textbf{2.74}\tabularnewline
\hline 
MM2 2 & 3.89 & 3.89 & 3.52 & \textbf{3.13}\tabularnewline
\hline 
MM2 3 & 2.48 & 2.47 & 2.45 & \textbf{2.41}\tabularnewline
\hline 
MM2 4 & 2.74 & 2.73 & 2.62 & \textbf{2.47}\tabularnewline
\hline 
MM2 5 & 2.90 & 2.82 & 2.67 & \textbf{2.64}\tabularnewline
\hline 
MM2 6 & 2.81 & 2.74 & 2.74 & \textbf{2.72}\tabularnewline
\hline 
MM2 7 & 2.73 & 2.72 & 2.76 & \textbf{2.67}\tabularnewline
\hline 
MM2 8 & 3.00 & 2.92 & 2.88 & \textbf{2.70}\tabularnewline
\hline 
MM2 9 & 3.27 & 3.21 & 2.95 & \textbf{2.79}\tabularnewline
\hline 
MM2 10 & 3.04 & 3.05 & 2.98 & \textbf{2.82}\tabularnewline
\hline 
its4s2 & 3.63 & 3.63 & 2.96 & \textbf{2.59}\tabularnewline
\hline 
its4s AGH & 3.15 & 3.05 & 2.77 & \textbf{2.64}\tabularnewline
\hline 
its4s NTIA & 2.94 & 2.91 & 2.53 & \textbf{2.38}\tabularnewline
\hline 
\end{tabular}
\par\end{centering}
\vspace{0.05in}
\caption{NBIC reported on the compared methods on public datasets. The NR and
AP (and AP2) methods produce identical results. (MOS: plain mean opinion
score; NR: Newton-Raphson; AP: Alternating Projection.)}
\label{table:bic}
\end{table}

\subsection{Confidence Interval of Quality Scores\label{subsec:Confidence-Interval-of}}

Table \ref{table:quality_ci} shows the average length of the CIs
on the compared methods on the 22 public datasets. The smaller the
number, the tighter the CI, thus more confident the estimation is.
For MOS, BT.500 and P.913, the CIs are calculated based on (\ref{eq:quality_ci_mos}).
For BT.500 and P.913, only the opinions scores after rejection are
taken into account. For the proposed methods NR and AP, the CIs are
calculated based on (\ref{eq:quality_ci_mle}). The proposed alternative
$CI_{2}$ (\ref{eq:quality_ci_mle_alt}) combined with AP is denoted
by AP2. It can be observed that the NR and AP yield the tightest CIs
compared to the other methods. AP2 also yields very tight CIs, except
for the NFLX public dataset, where the four outliers' very loose CIs
contributed to the loosening of the overall CIs (0.57). For some databases,
BT.500 generates wider confidence interval than the plain MOS. This
can be explained by the fact that subject rejection decreases the
number of samples, even though the variance may also be decreased.
Overall, the obtained confidence interval can be either narrower or
wider.

\begin{figure*}
\begin{centering}
\noindent\begin{minipage}[t]{1.3\columnwidth}%
\begin{center}
\includegraphics[width=1\columnwidth]{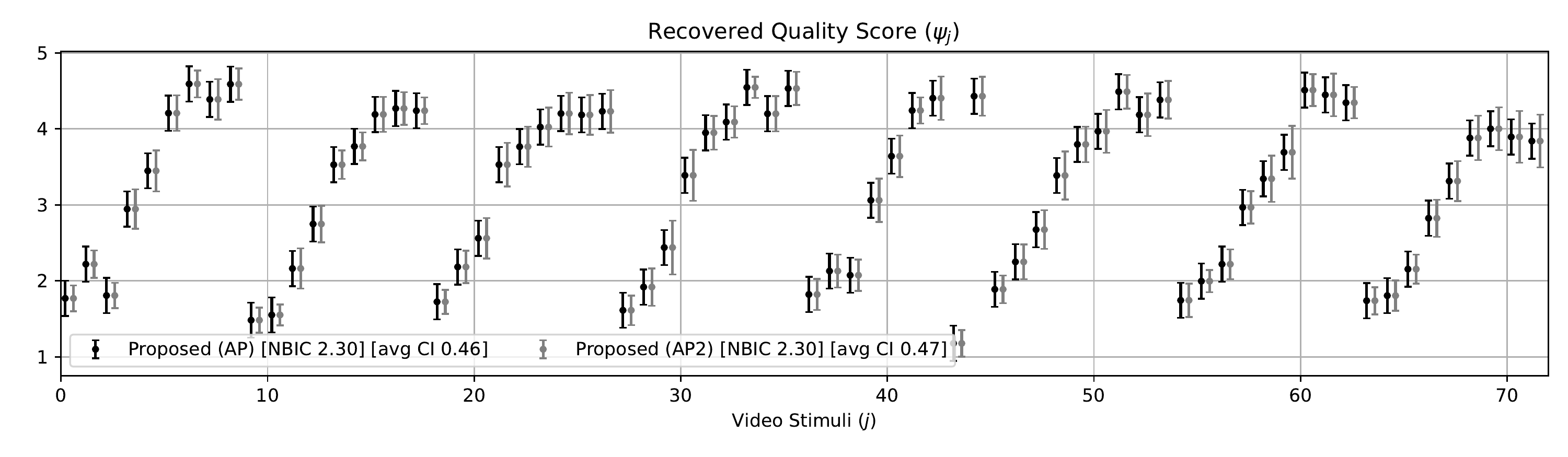}
\par\end{center}
\vspace{-0.2in}
\begin{center}
(a) VQEG HD3 dataset
\par\end{center}%
\end{minipage}
\par\end{centering}
\begin{centering}
\noindent\begin{minipage}[t]{1.3\columnwidth}%
\begin{center}
\includegraphics[width=1\columnwidth]{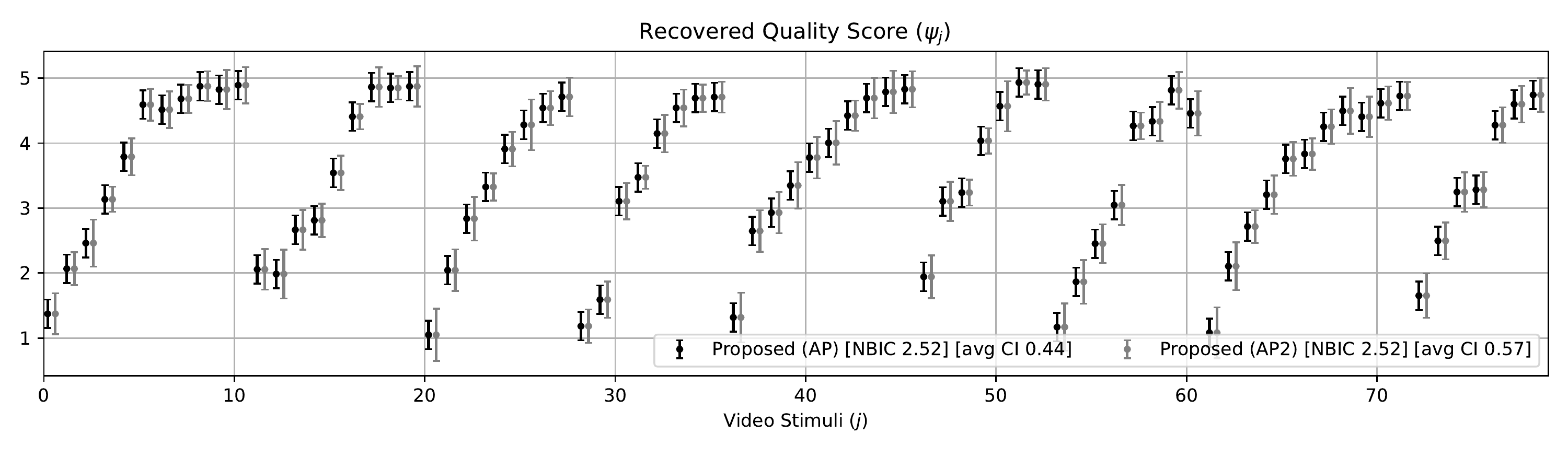}
\par\end{center}
\vspace{-0.2in}
\begin{center}
(b) NFLX Public dataset
\par\end{center}%
\end{minipage}
\par\end{centering}
\caption{\label{fig:quality-scores-ci-ap2}Recovered quality score $\psi_{j}$
and its CI for AP (CI based on (\ref{eq:quality_ci_mle})) and AP2
(CI based on (\ref{eq:quality_ci_mle_alt})) on (a) the VQEG HD3 dataset
and (b) the NFLX Public dataset. For each method compared, the NBIC
score (see Section \ref{subsec:Validation-using-Bayesian}) and the
average length of the confidence interval are reported. (avg CI: average
confidence interval; NBIC: Bayesian Information Criterion; AP: Alternating
Projection.)}
\end{figure*}

To visually compare the CIs for AP and AP2, we plot them in Figure
\ref{fig:quality-scores-ci-ap2}. It can be observed that AP yields
constant CIs across the stimuli whereas AP2 yields differentiated
CIs.

\begin{table}
\begin{centering}
\begin{tabular}{|c|cccc|}
\hline 
Dataset & MOS & BT.500 & P.913 & NR/AP (AP2)\tabularnewline
\hline 
VQEG HD3 & 0.59 & 0.60 & 0.49 & \textbf{0.46} (0.47)\tabularnewline
\hline 
NFLX Public & 0.62 & 0.54 & 0.5 & \textbf{0.44} (0.57)\tabularnewline
\hline 
HDTV Ph1 Exp1 & 0.50 & 0.61 & 0.48 & \textbf{0.46} (0.46)\tabularnewline
\hline 
HDTV Ph1 Exp2 & 0.57 & 0.57 & 0.53 & \textbf{0.48} (0.49)\tabularnewline
\hline 
HDTV Ph1 Exp3 & 0.56 & 0.59 & 0.52 & \textbf{0.48} (0.48)\tabularnewline
\hline 
HDTV Ph1 Exp4 & 0.63 & 0.63 & 0.52 & \textbf{0.47} (0.49)\tabularnewline
\hline 
HDTV Ph1 Exp5 & 0.57 & 0.57 & 0.53 & \textbf{0.49} (0.50)\tabularnewline
\hline 
HDTV Ph1 Exp6 & 0.50 & 0.51 & 0.48 & \textbf{0.45} (0.45)\tabularnewline
\hline 
ITU-T Supp23 Exp1 & 0.61 & 0.61 & 0.56 & \textbf{0.47} (0.50)\tabularnewline
\hline 
MM2 1 & 0.59 & 0.60 & 0.57 & \textbf{0.53} (0.55)\tabularnewline
\hline 
MM2 2 & 1.21 & 1.21 & 1.12 & \textbf{0.88} (0.99)\tabularnewline
\hline 
MM2 3 & 0.47 & 0.48 & 0.45 & \textbf{0.42} (0.43)\tabularnewline
\hline 
MM2 4 & 0.58 & 0.59 & 0.54 & \textbf{0.48} (0.51)\tabularnewline
\hline 
MM2 5 & 0.63 & 0.65 & 0.58 & \textbf{0.52} (0.56)\tabularnewline
\hline 
MM2 6 & 0.62 & 0.70 & 0.59 & \textbf{0.56} (0.57)\tabularnewline
\hline 
MM2 7 & 0.60 & 0.61 & 0.57 & \textbf{0.55} (0.55)\tabularnewline
\hline 
MM2 8 & 0.76 & 0.76 & 0.71 & \textbf{0.66} (0.68)\tabularnewline
\hline 
MM2 9 & 0.84 & 0.85 & 0.74 & \textbf{0.68} (0.71)\tabularnewline
\hline 
MM2 10 & 0.77 & 0.83 & 0.73 & \textbf{0.70} (0.70)\tabularnewline
\hline 
its4s2 & 0.82 & 0.82 & 0.66 & \textbf{0.60} (0.64)\tabularnewline
\hline 
its4s AGH & 0.68 & 0.68 & 0.61 & \textbf{0.56} (0.60)\tabularnewline
\hline 
its4s NTIA & 0.57 & 0.58 & 0.54 & \textbf{0.48} (0.50)\tabularnewline
\hline 
\end{tabular}
\par\end{centering}
\vspace{0.05in}
\caption{Average length of confidence intervals of the estimated quality scores
reported on the compared methods on public datasets. The NR and AP
methods are based on (\ref{eq:quality_ci_mle}) and they produce identical
results. AP2 is the AP method combined with the alternative CI calculation
(\ref{eq:quality_ci_mle_alt}). MOS: arithmetic mean of all opinion
scores; NR: Newton-Raphson; AP: Alternating Projection.}
\label{table:quality_ci}
\end{table}

\begin{figure}
\noindent\begin{minipage}[t]{1\columnwidth}%
\begin{center}
\includegraphics[width=1\columnwidth]{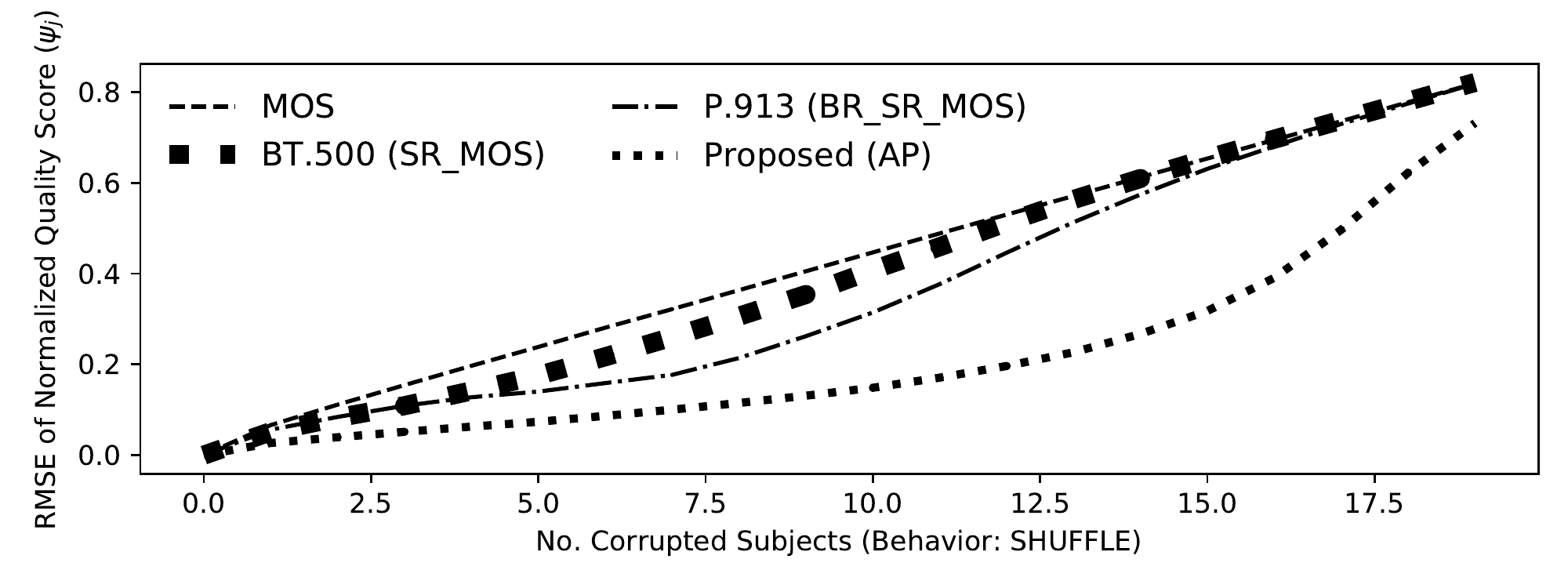}
\par\end{center}
\vspace{-0.2in}
\begin{center}
(a) VQEG HD3 dataset
\par\end{center}%
\end{minipage}
\begin{centering}
\par\end{centering}
\noindent\begin{minipage}[t]{1\columnwidth}%
\begin{center}
\includegraphics[width=1\columnwidth]{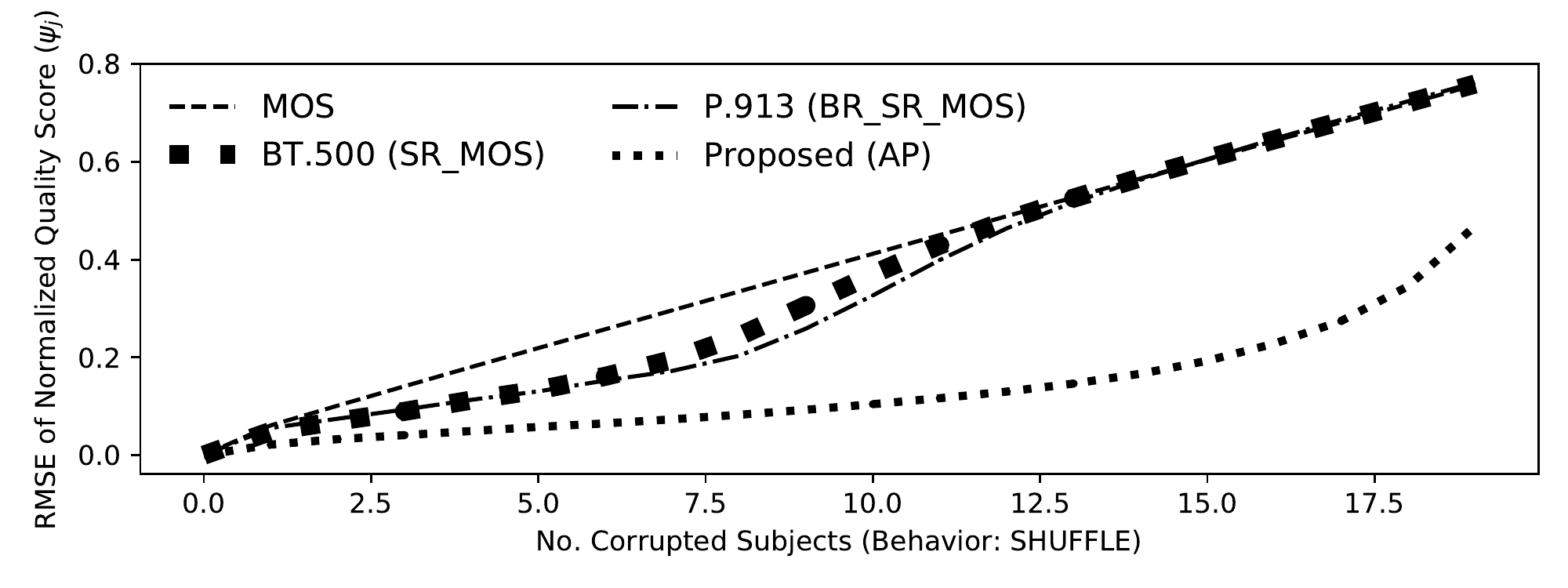}
\par\end{center}
\vspace{-0.2in}
\begin{center}
(b) NFLX Public dataset
\par\end{center}%
\end{minipage}
\caption{\label{fig:corruption-subj-grow}RMSE of the (normalized) recovered
quality score $\psi_{j}$ as a function of the number of corrupted
subjects, of the proposed method (AP) versus the other methods, of
(a) the VQEG HD3 dataset and (b) the NFLX Public dataset. The NR and
AP solvers yield identical results. The subject corruption is simulated,
in the way that the scores corresponding to a subject are scrambled.
The recovered quality score is normalized by subtracting the mean
and dividing by the standard deviation of the scores of the unaltered
dataset. (MOS: plain mean opinion score; SR: Subject Rejection; BR:
Bias Removal; AP: Alternating Projection.)}
\end{figure}

\begin{figure}
\noindent\begin{minipage}[t]{1\columnwidth}%
\begin{center}
\includegraphics[width=1\columnwidth]{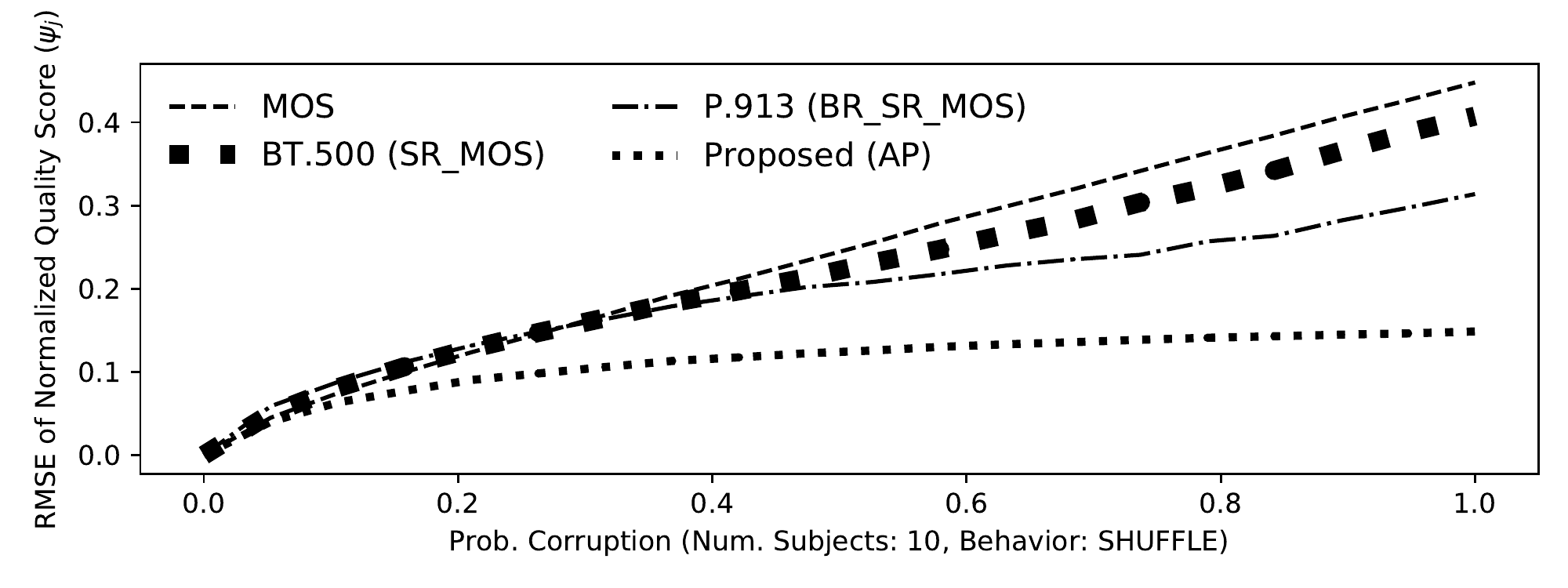}
\par\end{center}
\vspace{-0.2in}
\begin{center}
(a) VQEG HD3 dataset
\par\end{center}%
\end{minipage}
\begin{centering}
\par\end{centering}
\noindent\begin{minipage}[t]{1\columnwidth}%
\begin{center}
\includegraphics[width=1\columnwidth]{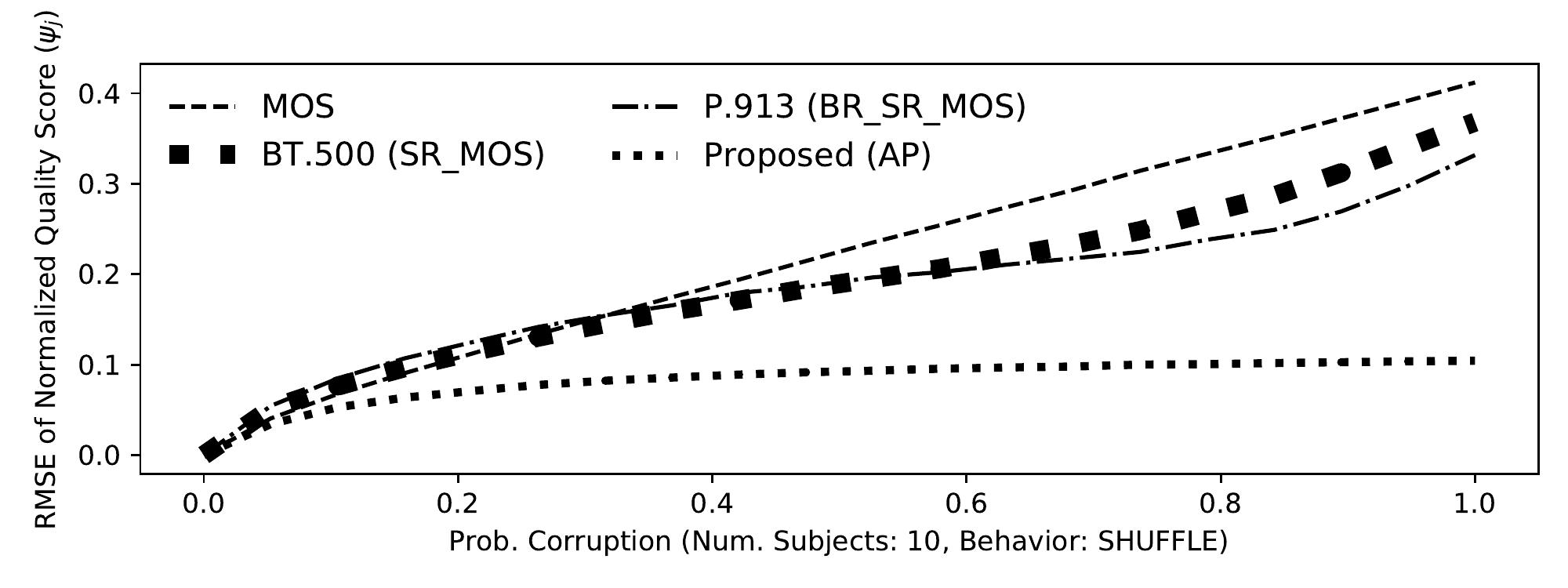}
\par\end{center}
\vspace{-0.2in}
\begin{center}
(b) NFLX Public dataset
\par\end{center}%
\end{minipage}
\caption{\label{fig:corruption-prob-grow}RMSE of the (normalized) recovered
quality score $\psi_{j}$ as a function of the probability of corruption
(fixing the number of corrupted subjects to 10), of the proposed method
(AP) versus the other methods, of (a) the VQEG HD3 dataset and (b)
the NFLX Public dataset. The NR and AP solvers yield identical results.
The subject corruption is simulated, in the way that the scores corresponding
to a subject are scrambled. The recovered quality score is normalized
by subtracting the mean and dividing by the standard deviation of
the scores of the unaltered dataset. (MOS: plain mean opinion score;
SR: Subject Rejection; BR: Bias Removal; AP: Alternating Projection.)}
\end{figure}

\subsection{Robustness against Outlier Subjects}

We demonstrate that the proposed method is much more effective in
dealing with (corrupted) outlier subjects compared to other methods.
We use the following methodology in our reporting of results. For
each method compared, we have a benchmark result, which is the recovered
quality scores obtained using \emph{that} method - for fairness -
on an \emph{unaltered full dataset} (note that for the NFLX Public
dataset, unlike the one used in Figure \ref{fig:Raw-opinion-scores},
\ref{fig:quality-scores} and \ref{fig:subject-scores}, we start
with the dataset where the corruption on the last four subjects has
been corrected). We then consider that a number of the subjects are
``corrupted'' and simulate it by randomly shuffling each corrupted
subject\textquoteright s votes among the video stimuli. We then run
each method compared on the partially corrupted datasets. The quality
scores recovered are normalized by subtracting the mean and dividing
by the standard deviation of the scores of the unaltered dataset.
The normalized scores are compared against the benchmark, and a root-mean-squared-error
(RMSE) value is reported.

\begin{figure*}
\begin{centering}
\noindent\begin{minipage}[t]{1.5\columnwidth}%
\begin{center}
\includegraphics[width=1\columnwidth]{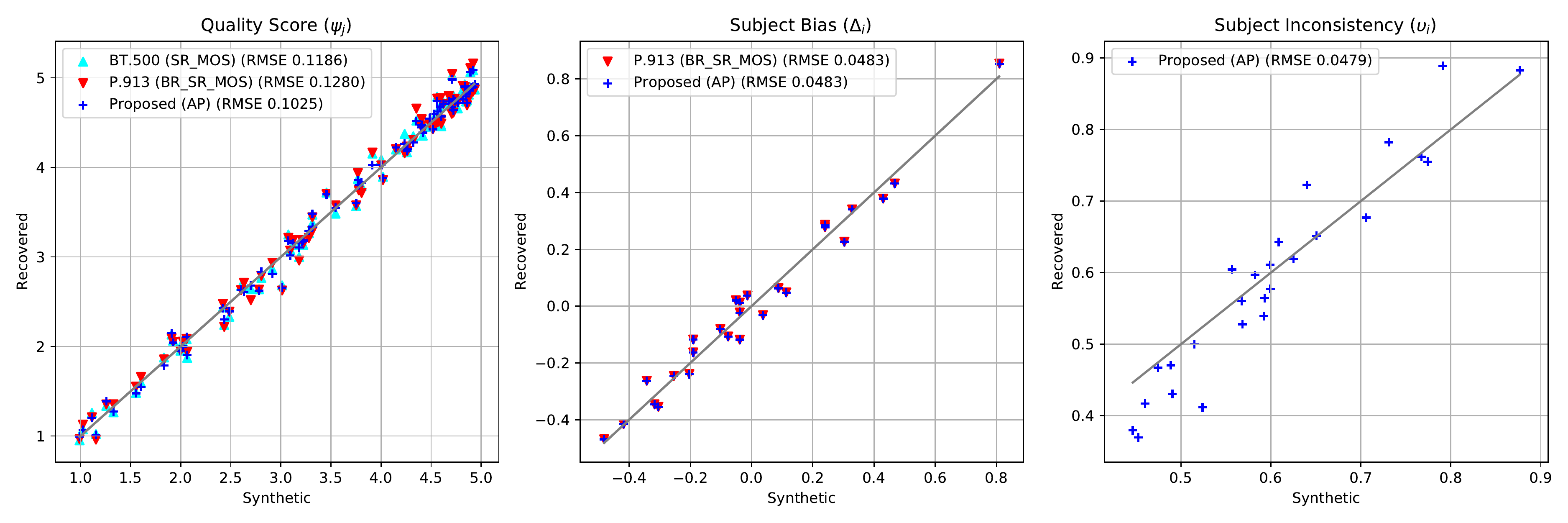}
\par\end{center}
\vspace{-0.2in}
\begin{center}
(a) Comparing BT.500, P.913 and AP
\par\end{center}%
\vspace{0.05in}
\end{minipage}
\noindent\begin{minipage}[t]{1.5\columnwidth}%
\begin{center}
\includegraphics[width=1\columnwidth]{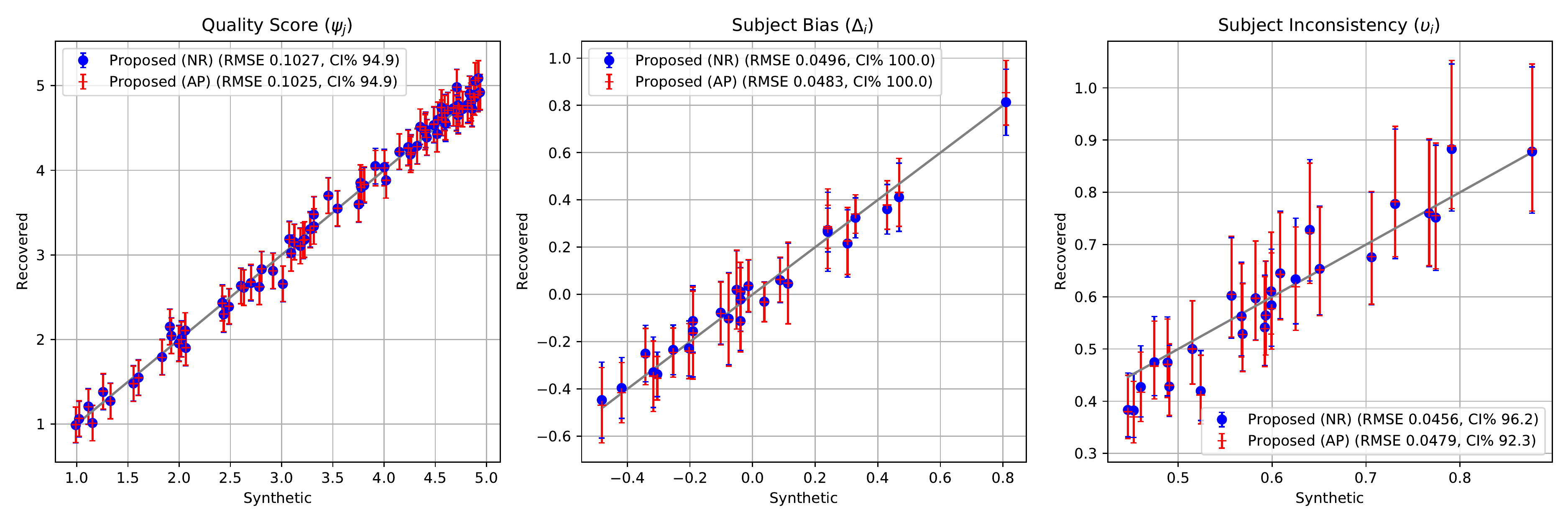}
\par\end{center}
\vspace{-0.2in}
\begin{center}
(b) Confidence Intervals of NR and AP
\par\end{center}%
\end{minipage}
\caption{\label{fig:synth_accuracy}Validation of the proposed NR and AP solvers
using synthetic data. The random samples are generated using the following
methodology. For each proposed solver, take the NFLX Public dataset,
run the solver to estimate the parameters. Treat the estimated parameters
as the ``synthetic'' parameters (or the ground truth), run simulations
to generate synthetic samples according to the model (\ref{eq:model}).
Run the solver again on the synthetic data to yield the ``recovered''
parameters. The x-axis shows the synthetic parameters and the y-axis
shows the recovered parameters. (a) Comparing the proposed AP with
BT.500 and P.913, (b) Proposed NR and AP with confidence intervals.
(NR: Newton-Raphson; AP: Alternating Projection.)\label{fig:synthetic-scatter}}
\end{centering}
\end{figure*}

Figure \ref{fig:corruption-subj-grow} reports the results on the
two datasets, comparing the proposed method with plain MOS, BT.500,
P.913 and the proposed NR and AP solvers, as the number of corrupted
subjects increases. It can be observed that in the presence of subject
corruption, The proposed method achieves a substantial gain over the
other methods. The reason is that the proposed model was able to capture
the variance of subjects explicitly and is able to compensate for
it. On the other hand, the other methods are only able to identify
part of the corrupted subjects. Meanwhile, traditional subject rejection
employs a set of hard coded parameters to determine outliers, which
may not be suited for all conditions. By contrast, the proposed model
naturally integrates the various subjective effects together and is
solved efficiently by the MLE formulation. In particular, the AP solver
can be interpreted as averaging the scores weighted by subject's consistency.
\emph{For corrupted subjects with large inconsistency scores, their
contribution to the final quality scores are limited.}

Figure \ref{fig:corruption-prob-grow} reports the results as we increase
the probability of corruption from 0 to 1 while fixing the number
of corrupted subjects to 10. It can be seen that as the corruption
probability increases, the RMSE increases linearly/near-linearly for
other methods, while the RMSE increases much slower for the proposed
method, and it saturates at a constant value without increasing further.
A simplified explanation is that, since only a subset of a subject\textquoteright s
scores is unreliable, discarding all of the subject\textquoteright s
scores is a waste of valuable subjective data, while the proposed
methods can effectively avoid that.

\subsection{Validations Using Synthetic Data\label{subsec:Validation-of-Solvers}}

Next, we demonstrate that the NR and AP solvers can accurately recover
the parameters of the proposed model. This is shown using synthetic
data, where the ground truth of the model parameters are known. In
this section, we considered only the NFLX Public dataset for simulations.
The random samples are generated using the following methodology.
For each proposed solver, we take the NFLX Public dataset and run
the solver to estimate the parameters. The parameters estimated from
a real dataset allow us to run simulations with practical settings.
We then treat the estimated parameters as the ``synthetic'' parameters
(hence the \emph{ground truth}), run simulations to generate synthetic
samples according to the model (\ref{eq:model}). Subsequently, we
run the solver again on the synthetic data to yield the ``recovered''
parameters.

\subsubsection{Validation of Solvers}

Figure \ref{fig:synthetic-scatter} shows the scatter plots of the
synthetic vs. recovered parameters, for the true quality $\psi_{j}$,
subjective bias $\Delta_{i}$ and subject inconsistency $\upsilon_{i}$
terms. It can be observed that the solvers recover the parameters
reasonably well. We have to keep in mind that the synthetic data,
differently from usual subjective scores of category rating, are continuous.
For discrete data, some specific problems would influence the obtained
results as described in \cite{janowski2019generalized}. Since those
problems are not the main topic of this paper we do not go into more
details and leave it as a future topic of research.

Figure \ref{fig:synthetic-scatter}(a) also shows the recovery result
of the BT.500 and P.913. It is noticeable that the recovered subject
biases by the AP method and the P.913 subject bias removal are very
similar. This should not be surprising, considering that the AP method
can be treated as a weighted and iterative generalization of the P.913
method.

\begin{table*}
\begin{centering}
\begin{tabular}{|c|c|ccc|ccc|}
\hline 
\multirow{2}{*}{Dataset} & MOS & \multicolumn{3}{c|}{NR} & \multicolumn{3}{c|}{AP (AP2)}\tabularnewline
\cline{2-8} \cline{3-8} \cline{4-8} \cline{5-8} \cline{6-8} \cline{7-8} \cline{8-8} 
 & $\psi_{j}$ & $\psi_{j}$ & $\Delta_{i}$ & $\upsilon_{i}$ & $\psi_{j}$ & $\Delta_{i}$ & $\upsilon_{i}$\tabularnewline
\hline 
VQEG HD3 & 93.3 & 93.6 & 93.9 & 93.0 & 93.2 (93.5) & 94.4 & 91.9\tabularnewline
\hline 
NFLX Public & 94.2 & 93.7 & 94.5 & 93.1 & 93.5 (97.5) & 94.1 & 92.3\tabularnewline
\hline 
HDTV Ph1 Exp1 & 93.9 & 94.1 & 93.9 & 93.1 & 93.8 (93.2) & 94.2 & 91.3\tabularnewline
\hline 
HDTV Ph1 Exp2 & 93.8 & 94.0 & 94.5 & 92.5 & 93.8 (94.1) & 94.0 & 91.2\tabularnewline
\hline 
HDTV Ph1 Exp3 & 93.9 & 93.9 & 94.4 & 92.5 & 93.7 (93.6) & 94.1 & 90.6\tabularnewline
\hline 
HDTV Ph1 Exp4 & 93.8 & 94.0 & 94.3 & 91.9 & 93.8 (94.1) & 94.1 & 90.9\tabularnewline
\hline 
HDTV Ph1 Exp5 & 93.8 & 94.1 & 94.2 & 92.2 & 93.9 (93.8) & 94.2 & 90.9\tabularnewline
\hline 
HDTV Ph1 Exp6 & 93.8 & 94.0 & 94.4 & 92.6 & 93.9 (93.6) & 94.0 & 91.0\tabularnewline
\hline 
ITU-T Supp23 Exp1 & 93.8 & 94.0 & 94.4 & 91.2 & 93.8 (94.5) & 94.9 & 90.0\tabularnewline
\hline 
MM2 1 & 93.5 & 92.8 & 95.4 & 92.6 & 92.5 (93.8) & 94.0 & 91.6\tabularnewline
\hline 
MM2 2 & 92.1 & 81.5 & 92.9 & 80.0 & 68.1 (87.5) & 92.1 & 75.4\tabularnewline
\hline 
MM2 3 & 94.4 & 93.6 & 95.1 & 93.4 & 93.4 (94.1) & 94.2 & 92.0\tabularnewline
\hline 
MM2 4 & 93.2 & 93.6 & 95.6 & 93.0 & 93.2 (94.7) & 95.1 & 92.0\tabularnewline
\hline 
MM2 5 & 93.2 & 93.2 & 95.7 & 92.7 & 91.8 (94.3) & 95.3 & 91.4\tabularnewline
\hline 
MM2 6 & 93.6 & 93.3 & 95.2 & 92.8 & 93.0 (93.8) & 94.1 & 91.4\tabularnewline
\hline 
MM2 7 & 93.6 & 93.3 & 95.2 & 92.8 & 92.9 (93.2) & 94.2 & 91.9\tabularnewline
\hline 
MM2 8 & 93.0 & 92.4 & 95.4 & 88.8 & 92.2 (92.6) & 94.5 & 87.0\tabularnewline
\hline 
MM2 9 & 93.2 & 93.3 & 94.8 & 89.1 & 92.8 (93.3) & 94.2 & 88.1\tabularnewline
\hline 
MM2 10 & 93.2 & 93.1 & 95.7 & 89.7 & 92.8 (92.3) & 94.5 & 87.9\tabularnewline
\hline 
its4s2 & 93.1 & 94.1 & 94.6 & 60.6 & 94.1 (94.0) & 94.2 & 59.2\tabularnewline
\hline 
its4s AGH & 93.6 & 94.0 & 94.4 & 90.4 & 94.0 (94.8) & 94.4 & 89.7\tabularnewline
\hline 
its4s NTIA & 93.9 & 94.4 & 94.7 & 86.1 & 94.3 (95.0) & 95.1 & 85.6\tabularnewline
\hline 
\end{tabular}
\par\end{centering}
\vspace{0.05in}
\caption{Average confidence interval coverage (CI\%) reported on public datasets.
For each proposed solver and each dataset, run the solver to estimate
the parameters. Treat the estimated parameters as ``synthetic''
parameters, run simulations to generate synthetic samples according
to the model (\ref{eq:model}) (except for MOS, whose samples are
generated according to (\ref{eq:model_mos})). Run the solver again
on the synthetic data to yield the ``recovered'' parameters and
their confidence intervals. The reported ``CI\%'' is the percentage
of occurrences when the synthetic ground truth falls within the confidence
interval. For each dataset, the simulation is run 100 times with different
seeds. Note that for both MOS and the proposed NR and AP methods,
the CI\% is slightly below 95\%, due to the underlying Gaussian assumption
used instead of the legitimate Student's \emph{t}-distribution. (MOS:
plain mean opinion score; NR: Newton-Raphson; AP: Alternating Projection.)
For $\psi_{j}$ of AP, two CI\% are reported: the CI based on (\ref{eq:quality_ci_mle})
and (in the parenthesis) the alternative CI calculated based on (\ref{eq:quality_ci_mle_alt}).}
\label{table:synth_ci_coverage}
\end{table*}

\begin{table*}
\begin{centering}
\begin{tabular}{|c|ccccc|cc|}
\hline 
\multirow{2}{*}{Dataset} & \multicolumn{5}{c|}{Mean Runtime (seconds)} & \multicolumn{2}{c|}{No. Iterations}\tabularnewline
\cline{2-8} \cline{3-8} \cline{4-8} \cline{5-8} \cline{6-8} \cline{7-8} \cline{8-8} 
 & MOS & BT.500 & P.913 & NR & AP & NR & AP\tabularnewline
\hline 
VQEG HD3 & 5.2e-4 & 1.5e-2 & 1.5e-2 & 2.1e-1 & 4.3e-3 & 26.2 & 12.1\tabularnewline
\hline 
NFLX Public & 5.7e-4 & 1.8e-2 & 1.9e-2 & 2.8e-1 & 4.5e-3 & 34.5 & 11.8\tabularnewline
\hline 
HDTV Ph1 Exp1 & 7.7e-4 & 3.3e-2 & 3.4e-2 & 2.0e-1 & 4.6e-3 & 23.4 & 10.3\tabularnewline
\hline 
HDTV Ph1 Exp2 & 7.8e-4 & 3.3e-2 & 3.4e-2 & 2.8e-1 & 4.9e-3 & 33.2 & 11.3\tabularnewline
\hline 
HDTV Ph1 Exp3 & 7.8e-4 & 3.3e-2 & 3.4e-2 & 2.5e-1 & 4.7e-3 & 29.4 & 10.7\tabularnewline
\hline 
HDTV Ph1 Exp4 & 7.6e-4 & 3.3e-2 & 3.4e-2 & 3.3e-1 & 5.0e-3 & 38.3 & 11.5\tabularnewline
\hline 
HDTV Ph1 Exp5 & 7.8e-4 & 3.3e-2 & 3.4e-2 & 2.7e-1 & 4.7e-3 & 31.3 & 10.8\tabularnewline
\hline 
HDTV Ph1 Exp6 & 7.6e-4 & 3.3e-2 & 3.4e-2 & 2.2e-1 & 4.6e-3 & 25.8 & 10.7\tabularnewline
\hline 
ITU-T Supp23 Exp1 & 8.1e-4 & 3.5e-2 & 3.5e-2 & 3.4e-1 & 5.0e-3 & 36.0 & 11.6\tabularnewline
\hline 
MM2 1 & 4.9e-4 & 1.3e-2 & 1.3e-2 & 2.1e-1 & 4.3e-3 & 27.4 & 12.4\tabularnewline
\hline 
MM2 2 & 4.0e-4 & 1.0e-2 & 1.1e-2 & 5.8e-1 & 1.4e-2 & 78.0 & 54.9\tabularnewline
\hline 
MM2 3 & 5.3e-4 & 1.3e-2 & 1.4e-2 & 1.8e-1 & 4.2e-3 & 23.3 & 11.6\tabularnewline
\hline 
MM2 4 & 5.0e-4 & 1.3e-2 & 1.4e-2 & 2.6e-1 & 4.6e-3 & 33.4 & 13.8\tabularnewline
\hline 
MM2 5 & 5.0e-4 & 1.3e-2 & 1.4e-2 & 2.9e-1 & 6.0e-3 & 37.3 & 19.3\tabularnewline
\hline 
MM2 6 & 4.8e-4 & 1.2e-2 & 1.3e-2 & 2.2e-1 & 4.3e-3 & 28.8 & 13.1\tabularnewline
\hline 
MM2 7 & 4.8e-4 & 1.2e-2 & 1.3e-2 & 2.0e-1 & 4.2e-3 & 25.6 & 12.3\tabularnewline
\hline 
MM2 8 & 4.3e-4 & 1.1e-2 & 1.1e-2 & 2.7e-1 & 5.5e-3 & 35.3 & 18.7\tabularnewline
\hline 
MM2 9 & 4.3e-4 & 1.1e-2 & 1.2e-2 & 2.8e-1 & 5.1e-3 & 36.5 & 16.8\tabularnewline
\hline 
MM2 10 & 4.3e-4 & 1.1e-2 & 1.2e-2 & 2.3e-1 & 4.8e-3 & 29.8 & 15.4\tabularnewline
\hline 
its4s2 & 3.3e-3 & 2.5e-1 & 2.5e-1 & 1.1e+0 & 1.3e-2 & 49.8 & 13.3\tabularnewline
\hline 
its4s AGH & 8.7e-4 & 4.1e-2 & 4.2e-2 & 3.5e-1 & 5.3e-3 & 39.4 & 11.6\tabularnewline
\hline 
its4s NTIA & 2.6e-3 & 1.6e-1 & 1.6e-1 & 6.4e-1 & 1.1e-2 & 46.2 & 11.3\tabularnewline
\hline 
\end{tabular}
\par\end{centering}
\vspace{0.05in}
\caption{Average runtime in seconds and number of iterations (for NR and AP)
reported on public datasets. AP2 has the same runtime as AP. For each
proposed solver and each dataset, run the solver to estimate the parameters.
Treat the estimated parameters and the ``synthetic'' parameters,
run simulations to generate synthetic samples according to the model
(\ref{eq:model}) (except for MOS, whose samples are generated according
to (\ref{eq:model_mos})). Run the solver again on the synthetic data.
For each dataset, the simulation is run 100 times with different seeds,
and the mean is reported. For NR and AP, also reported are the number
of iterations. (MOS: plain mean opinion score; NR: Newton-Raphson;
AP: Alternating Projection.)}
\label{table:synth_runtime}
\end{table*}

\begin{figure*}
\begin{centering}
\noindent\begin{minipage}[t]{1\columnwidth}%
\begin{center}
\includegraphics[width=1\columnwidth]{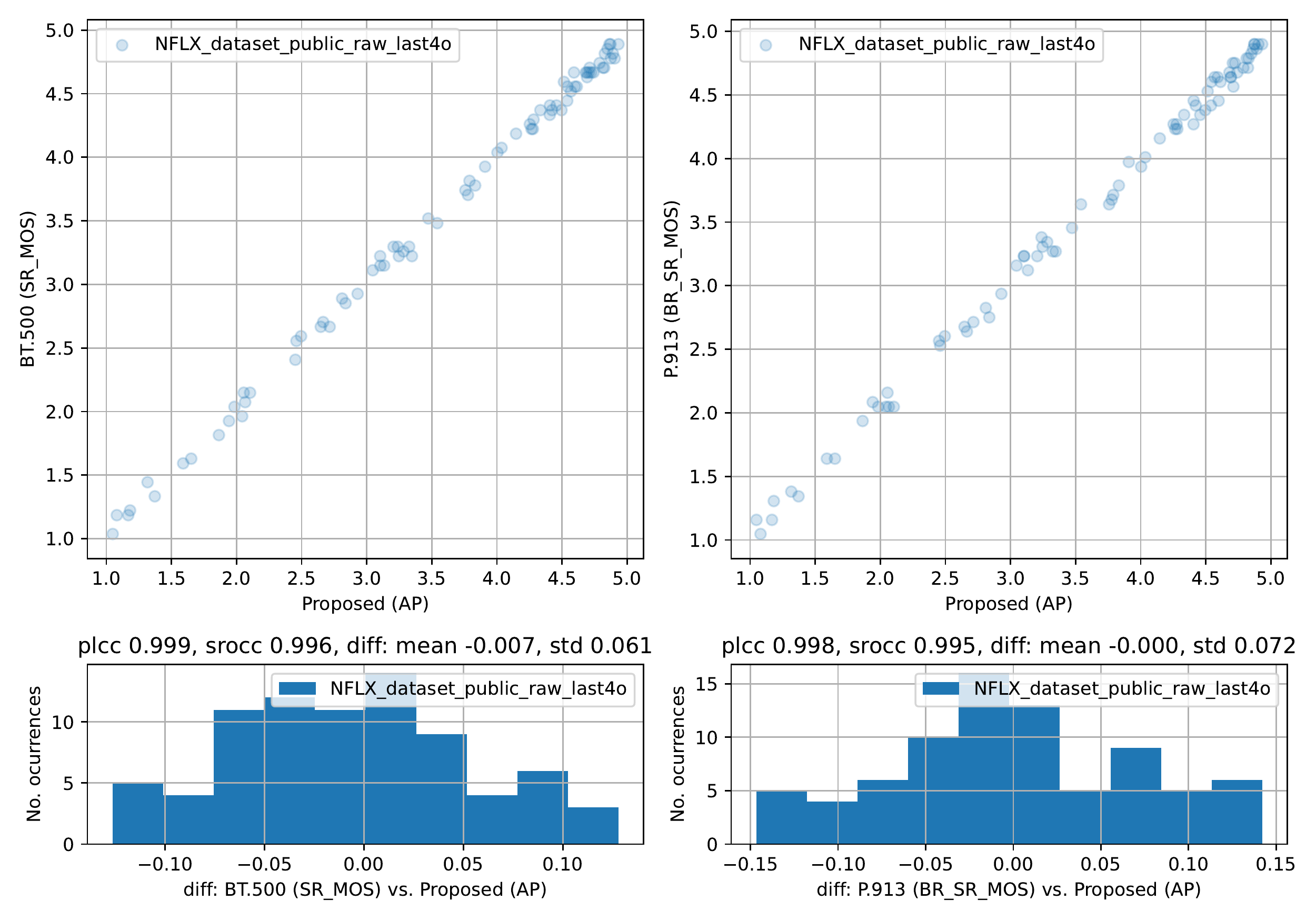}
\par\end{center}
\vspace{-0.1in}
\begin{center}
(a) NFLX Public dataset (lab test)
\par\end{center}%
\end{minipage}
\noindent\begin{minipage}[t]{1\columnwidth}%
\begin{center}
\includegraphics[width=1\columnwidth]{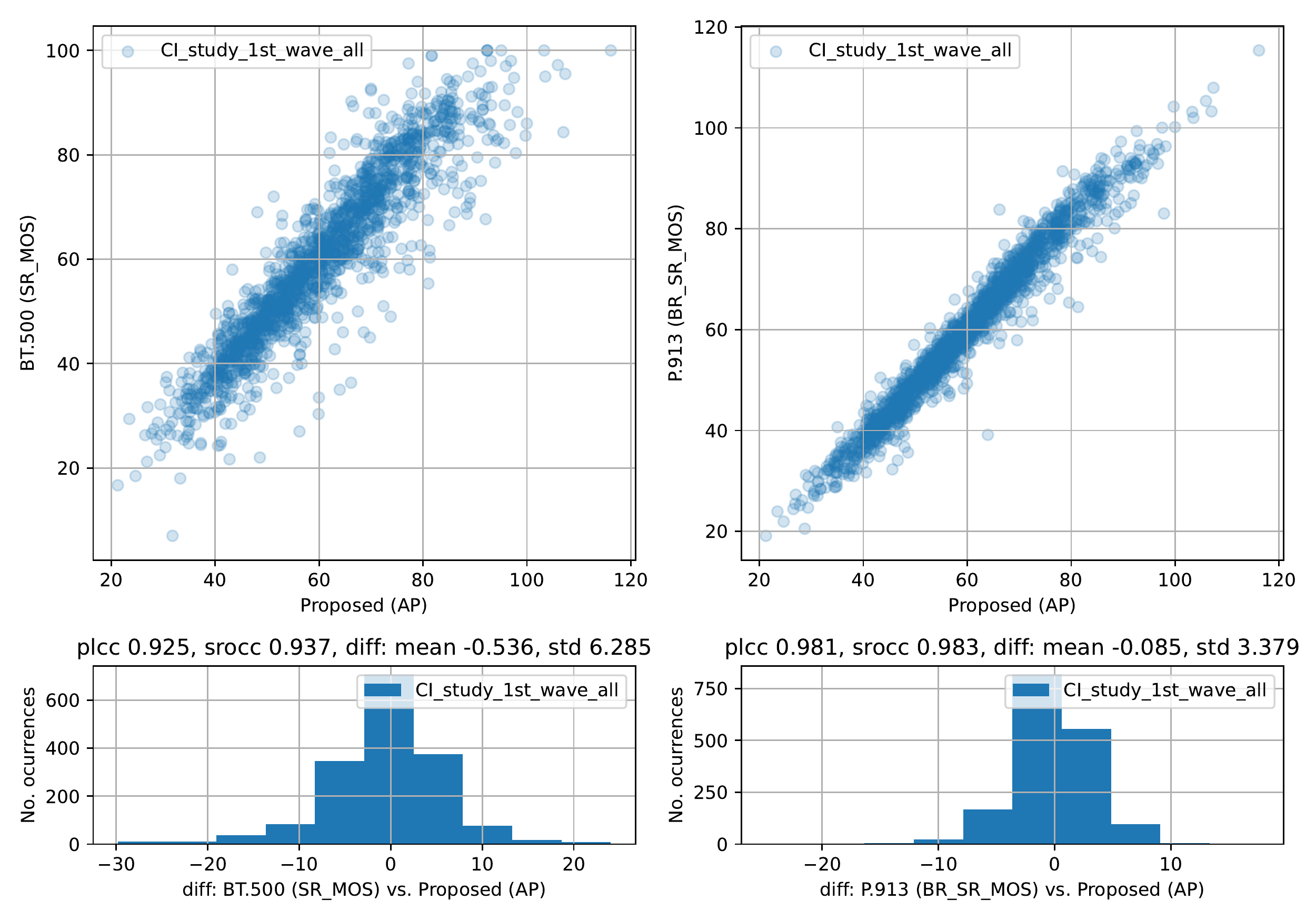}
\par\end{center}
\vspace{-0.1in}
\begin{center}
(b) NFLX Crowdsourcing 1st Wave dataset
\par\end{center}%
\end{minipage}
\par\end{centering}
\caption{\label{fig:scatter_bt500_p913_vs_ap}Scatter plots of recovered quality
scores $\psi_{j}$ between BT.500 and AP (left) and between P.913
and AP (right) for (a) the NFLX Public dataset and (b) NFLX Crowdsourcing
1st Wave dataset. The diagram below shows the histogram of the difference
values. The Pearson correlation (PLCC), Spearman correlation (SROCC),
the mean and the standard deviation of the difference values are also
reported. The NR and AP solvers yield identical results. (MOS: plain
mean opinion score; SR: Subject Rejection; BR: Bias Removal; AP: Alternating
Projection.)}
\end{figure*}

\subsubsection{Validation of Confidence Intervals}

Also plotted in Figure \ref{fig:synthetic-scatter}(b) are the confidence
intervals of the recovered parameters. The reported ``CI\%'' is
the percentage of occurrences where the synthetic ground truth falls
within the confidence interval. By definition, we expect the CI\%
to be 95\% on average. To verify this, we run the same simulation
on the 22 public datasets. For each dataset, the simulation is run
100 times with different seeds. The result is shown in Table \ref{table:synth_ci_coverage}.
We compare the proposed NR and AP methods with the plain MOS. It can
be seen that all methods yield CI\% to be very close to 95\%, but
slightly below. The explanation is that both have assumed that the
underlying distribution is Gaussian, but with both the mean and standard
deviation unknown, one should use a Student's \emph{t}-distribution
instead. If the \emph{t}-distribution is used, the coefficient can
no longer be a fixed value 1.96 but is a function of the number of
subjects and repetitions.

\subsection{Runtime and Iterations}

We then evaluate the runtime of the proposed NR and AP methods compared
to the others. The results of 100 simulations runs (based on the similar
methodology as in the previous sections) of each methods are reported
in Table \ref{table:synth_runtime}. The results reveal the order
of magnitude of the algorithms compared. The plain MOS is typically
the fastest, while the BT.500 and P.913 are two magnitude slower.
The NR and AP algorithms are three and one magnitude slower, respectively.
Noteably, the AP runs faster than BT.500 and P.913, and is about 50x
faster than the NR. The AP also requires about half the number of
iterations to reach convergence than the NR.

\subsection{Consistency Under Challenging Conditions}

In this section, we demonstrate that the proposed methods are the
most valuable when the test conditions are challenging. Traditional
lab tests are typically conducted in a highly controlled environment.
When analyzing the different methods using these lab datasets, we
find that the recovered quality scores using BT.500, P.913 and the
proposed AP (and NR) method are highly consistent. However, when the
test conditions are challenging, for example, in a crowdsourced test,
the different methods could yield quite inconsistent results. This
is illustrated in Figure \ref{fig:scatter_bt500_p913_vs_ap}, where
the quality scores recovered by different methods are compared in
scatter plots. Figure \ref{fig:scatter_bt500_p913_vs_ap} (a) is based
on the NFLX Public dataset (a lab test), and Figure \ref{fig:scatter_bt500_p913_vs_ap}
(b) is based on a crowdsourced test conducted by Netflix (details
to follow). It can be seen that the results from the crowdsourced
test are much less consistent across methods.

Which method yields the most accurate recovery? Unfortunately, we
cannot directly answer this question since we do not have the ground
truth. However, we can demonstrate that \emph{the proposed AP method
can yield more consistency (or less variability)}, and thus is more
desirable. We demonstrate this using 1) the crowdsourced test where
we correlate the result using all the data with the one using only
10\% of the data, and 2) a cross-lab test, where the same stimuli
are tested at different lab locations. 

\subsubsection{NFLX Crowdsourcing Test}

\begin{figure*}
\begin{center}
\noindent\begin{minipage}[t]{1.5\columnwidth}%
\includegraphics[width=1\columnwidth]{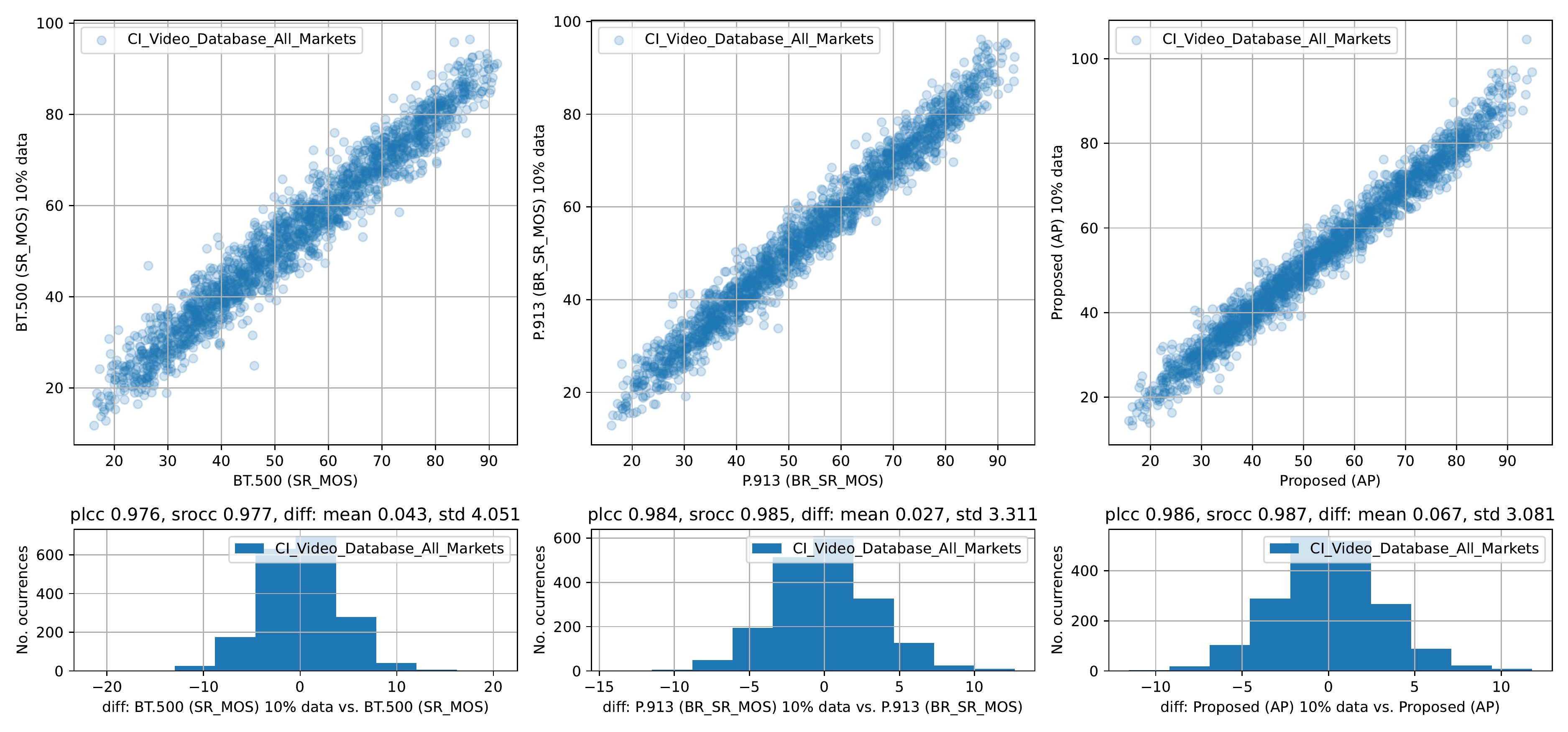}
\end{minipage}
\par\end{center}%
\caption{\label{fig:scatter_full_vs_10perc}Scatter plots of recovered quality
scores $\psi_{j}$ between using full data (x-axis) vs. using only
10\% of the data (randomly sampled, y-axis) of the NFLX Crowdsourcing
2nd Wave dataset. The left, middle and right plots are based on BT.500,
P.913 and the proposed AP, respectively. The diagram below shows the
histogram of the difference values. The Pearson correlation (PLCC),
Spearman correlation (SROCC), the mean and the standard deviation
of the difference values are also reported. The NR and AP solvers
yield identical results. (MOS: plain mean opinion score; SR: Subject
Rejection; BR: Bias Removal; AP: Alternating Projection.)}
\end{figure*}

The NFLX Crowdsourcing Test was conducted on 154 video contents and
1859 video stimuli, each 10-second long. It consists of two datasets:
1) the 1st Wave dataset, a small pre-test with 20 votes per stimulus
on average, and 2) the 2nd Wave dataset, the full test with 290 votes
per stimulus on average. In this section, we use the 2nd Wave dataset.
For each of the methods evaluated, we compare the quality scores recovered
using the full dataset with the scores recovered using only 10\% of
the data, randomly sampled. The scatter plots are shown in Figure
\ref{fig:scatter_full_vs_10perc}, which also reports the correlations
and the mean and standard deviation of the difference values. It can
be seen that the proposed AP method yields the best consistency and
the least variability among the three methods. In addition, P.913
does better than BT.500, due to the removal of the subject bias.

\begin{table*}
\begin{centering}
\noindent\begin{minipage}[t]{1.5\columnwidth}%
\begin{minipage}[t]{0.33\columnwidth}%
\begin{center}
\tiny%
\begin{tabular}{|c|c|c|c|c|}
\hline 
Lab & 1 & 4 & 6 & 8\tabularnewline
\hline 
\hline 
1 & 1.0 & 0.944 & 0.943 & 0.948\tabularnewline
\hline 
4 &  & 1.0 & \textbf{\textbf{0.957}} & 0.941\tabularnewline
\hline 
6 &  &  & 1.0 & 0.944\tabularnewline
\hline 
8 &  &  &  & 1.0\tabularnewline
\hline 
\end{tabular}
\par\end{center}
\begin{center}
BT.500
\par\end{center}%
\end{minipage}%
\begin{minipage}[t]{0.33\columnwidth}%
\begin{center}
\tiny%
\begin{tabular}{|c|c|c|c|c|}
\hline 
Lab & 1 & 4 & 6 & 8\tabularnewline
\hline 
\hline 
1 & 1.0 & 0.950 & 0.942 & 0.95\tabularnewline
\hline 
4 &  & 1.0 & 0.9556 & 0.940\tabularnewline
\hline 
6 &  &  & 1.0 & 0.944\tabularnewline
\hline 
8 &  &  &  & 1.0\tabularnewline
\hline 
\end{tabular}
\par\end{center}
\begin{center}
P.913
\par\end{center}%
\end{minipage}%
\begin{minipage}[t]{0.33\columnwidth}%
\begin{center}
\tiny%
\begin{tabular}{|c|c|c|c|c|}
\hline 
Lab & 1 & 4 & 6 & 8\tabularnewline
\hline 
\hline 
1 & 1.0 & \textbf{\textbf{0.952}} & \textbf{\textbf{0.949}} & \textbf{\textbf{0.958}}\tabularnewline
\hline 
4 &  & 1.0 & 0.957 & \textbf{\textbf{0.945}}\tabularnewline
\hline 
6 &  &  & 1.0 & \textbf{\textbf{0.948}}\tabularnewline
\hline 
8 &  &  &  & 1.0\tabularnewline
\hline 
\end{tabular}
\par\end{center}
\begin{center}
Proposed AP
\par\end{center}%
\end{minipage}
\begin{center}
(a) 525 Line Low
\par\end{center}%
\vspace{0.05in}
\end{minipage}

\noindent\begin{minipage}[t]{1.5\columnwidth}%
\begin{minipage}[t]{0.33\columnwidth}%
\begin{center}
\tiny%
\begin{tabular}{|c|c|c|c|c|}
\hline 
Lab & 1 & 4 & 6 & 8\tabularnewline
\hline 
\hline 
1 & 1.0 & 0.890 & 0.900 & 0.915\tabularnewline
\hline 
4 &  & 1.0 & \textbf{\textbf{0.881}} & \textbf{\textbf{0.850}}\tabularnewline
\hline 
6 &  &  & 1.0 & \textbf{\textbf{0.875}}\tabularnewline
\hline 
8 &  &  &  & 1.0\tabularnewline
\hline 
\end{tabular}
\par\end{center}
\begin{center}
BT.500
\par\end{center}%
\end{minipage}%
\begin{minipage}[t]{0.33\columnwidth}%
\begin{center}
\tiny%
\begin{tabular}{|c|c|c|c|c|}
\hline 
Lab & 1 & 4 & 6 & 8\tabularnewline
\hline 
\hline 
1 & 1.0 & 0.888 & 0.903 & 0.906\tabularnewline
\hline 
4 &  & 1.0 & 0.867 & 0.834\tabularnewline
\hline 
6 &  &  & 1.0 & 0.874\tabularnewline
\hline 
8 &  &  &  & 1.0\tabularnewline
\hline 
\end{tabular}
\par\end{center}
\begin{center}
P.913
\par\end{center}%
\end{minipage}%
\begin{minipage}[t]{0.33\columnwidth}%
\begin{center}
\tiny%
\begin{tabular}{|c|c|c|c|c|}
\hline 
Lab & 1 & 4 & 6 & 8\tabularnewline
\hline 
\hline 
1 & 1.0 & \textbf{\textbf{0.906}} & \textbf{\textbf{0.911}} & \textbf{\textbf{0.915}}\tabularnewline
\hline 
4 &  & 1.0 & 0.876 & 0.823\tabularnewline
\hline 
6 &  &  & 1.0 & 0.833\tabularnewline
\hline 
8 &  &  &  & 1.0\tabularnewline
\hline 
\end{tabular}
\par\end{center}
\begin{center}
Proposed AP
\par\end{center}%
\end{minipage}
\begin{center}
(b) 525 Line High
\par\end{center}%
\vspace{0.05in}
\end{minipage}

\noindent\begin{minipage}[t]{1.5\columnwidth}%
\begin{minipage}[t]{0.33\columnwidth}%
\begin{center}
\tiny%
\begin{tabular}{|c|c|c|c|c|}
\hline 
Lab & 2 & 3 & 5 & 7\tabularnewline
\hline 
\hline 
2 & 1.0 & 0.743 & 0.913 & 0.914\tabularnewline
\hline 
3 &  & 1.0 & 0.812 & 0.705\tabularnewline
\hline 
5 &  &  & 1.0 & 0.900\tabularnewline
\hline 
7 &  &  &  & 1.0\tabularnewline
\hline 
\end{tabular}
\par\end{center}
\begin{center}
BT.500
\par\end{center}%
\end{minipage}%
\begin{minipage}[t]{0.33\columnwidth}%
\begin{center}
\tiny%
\begin{tabular}{|c|c|c|c|c|}
\hline 
Lab & 2 & 3 & 5 & 7\tabularnewline
\hline 
\hline 
2 & 1.0 & 0.764 & 0.908 & 0.904\tabularnewline
\hline 
3 &  & 1.0 & 0.840 & 0.755\tabularnewline
\hline 
5 &  &  & 1.0 & 0.905\tabularnewline
\hline 
7 &  &  &  & 1.0\tabularnewline
\hline 
\end{tabular}
\par\end{center}
\begin{center}
P.913
\par\end{center}%
\end{minipage}%
\begin{minipage}[t]{0.33\columnwidth}%
\begin{center}
\tiny%
\begin{tabular}{|c|c|c|c|c|}
\hline 
Lab & 2 & 3 & 5 & 7\tabularnewline
\hline 
\hline 
2 & 1.0 & \textbf{\textbf{0.814}} & \textbf{\textbf{0.926}} & \textbf{\textbf{0.923}}\tabularnewline
\hline 
3 &  & 1.0 & \textbf{\textbf{0.875 }} & \textbf{\textbf{0.804}}\tabularnewline
\hline 
5 &  &  & 1.0 & \textbf{\textbf{0.918}}\tabularnewline
\hline 
7 &  &  &  & 1.0\tabularnewline
\hline 
\end{tabular}
\par\end{center}
\begin{center}
Proposed AP
\par\end{center}%
\end{minipage}
\begin{center}
(c) 625 Line Low
\par\end{center}%
\vspace{0.05in}
\end{minipage}

\noindent\begin{minipage}[t]{1.5\columnwidth}%
\begin{minipage}[t]{0.33\columnwidth}%
\begin{center}
\tiny%
\begin{tabular}{|c|c|c|c|c|}
\hline 
Lab & 2 & 3 & 5 & 7\tabularnewline
\hline 
\hline 
2 & 1.0 & 0.790 & \textbf{\textbf{0.853}} & \textbf{\textbf{0.818}}\tabularnewline
\hline 
3 &  & 1.0 & 0.818  & 0.836\tabularnewline
\hline 
5 &  &  & 1.0 & 0.869\tabularnewline
\hline 
7 &  &  &  & 1.0\tabularnewline
\hline 
\end{tabular}
\par\end{center}
\begin{center}
BT.500
\par\end{center}%
\end{minipage}%
\begin{minipage}[t]{0.33\columnwidth}%
\begin{center}
\tiny%
\begin{tabular}{|c|c|c|c|c|}
\hline 
Lab & 2 & 3 & 5 & 7\tabularnewline
\hline 
\hline 
2 & 1.0 & 0.764 & 0.794 & 0.737\tabularnewline
\hline 
3 &  & 1.0 & \textbf{\textbf{0.826}} & 0.834\tabularnewline
\hline 
5 &  &  & 1.0 & 0.849\tabularnewline
\hline 
7 &  &  &  & 1.0\tabularnewline
\hline 
\end{tabular}
\par\end{center}
\begin{center}
P.913
\par\end{center}%
\end{minipage}%
\begin{minipage}[t]{0.33\columnwidth}%
\begin{center}
\tiny%
\begin{tabular}{|c|c|c|c|c|}
\hline 
Lab & 2 & 3 & 5 & 7\tabularnewline
\hline 
\hline 
2 & 1.0 & \textbf{\textbf{0.829}} & 0.818 & 0.800\tabularnewline
\hline 
3 &  & 1.0 & 0.825 & \textbf{\textbf{0.860}}\tabularnewline
\hline 
5 &  &  & 1.0 & \textbf{\textbf{0.874}}\tabularnewline
\hline 
7 &  &  &  & 1.0\tabularnewline
\hline 
\end{tabular}
\par\end{center}
\begin{center}
Proposed AP
\par\end{center}%
\end{minipage}
\begin{center}
(d) 625 Line High
\par\end{center}%
\vspace{0.05in}
\end{minipage}
%\vspace{0.05in}
\caption{Cross-lab Pearson correlation coefficient (PLCC) results for the VQEG
FRTV Phase I study. The study produces four datasets: 1) 525 Line
Low, 2) 525 Line High, 3) 625 Line Low and 4) 625 Line High. In total
8 labs participated in the study, and each dataset is evaluated by
4 of the 8 labs. The quality scores are recovered by three methods:
BT.500, P.913 and the proposed AP. The best result among the three
methods is in bold.}
\label{table:cross_lab}
\end{centering}
\end{table*}

\subsubsection{VQEG FRTV Phase I Study}

The VQEG full-reference television (FRTV) Phase I study \cite{frtvp1}
examines full-reference objective quality models that predicted the
quality of standard definition television (625-line and 525-line).
The study produces four datasets: 1) 525 Line Low, 2) 525 Line High,
3) 625 Line Low and 4) 625 Line High. In total 8 labs participated
in the study, and each dataset is evaluated by 4 of the 8 labs. Table
\ref{table:cross_lab} shows the cross-lab Pearson correlation coefficient.
The quality scores are recovered by three methods: BT.500, P.913 and
the proposed AP. From the result, one can conclude that among the
three methods, statistically, the proposed AP method yields the best
consistency across labs.

\section{Conclusions}

In the paper, we proposed a simple model to account for two of the
most dominant effects of test subject inaccuracy: subject bias and
subject inconsistency. We further proposed to solve the model parameters
through maximum likelihood estimation and presented two numerical
solvers. We compared the proposed methodology with the standardized
recommendations including ITU-R BT.500 and ITU-T P.913, and showed
that the proposed methods are the most valuable when the test conditions
are challenging (for example, crowdsourcing and cross-lab studies),
offering advantages such as better model-data fit, tighter confidence
intervals, better robustness against subject outliers, the absence
of hard coded parameters and thresholds, and auxiliary information
on test subjects. We believe the proposed methodology is generally
suitable for subjective evaluation of perceptual audiovisual quality
in multimedia and television services, and we propose to update the
corresponding recommendations with the methods presented.

% or


\appendices

\section{Mathematical Descriptions of BT.500 and P.913\label{sec:Appendix:-Mathematical-Descripti}}

\subsection{BT.500 Subject Rejection}

In this section, we give mathematical descriptions of the subject
rejection method standardized in ITU-R BT.500-14 and the subject bias
removal method in ITU-T P.913. Let $u_{ijr}$ be the opinion score
voted by subject $i$ on stimulus $j$ in repetition $r$. Note that,
in BT.500-14, the notation $j$ is used to indicate test condition
and $k$ is used to indicate sequence/image; in this paper, the test
condition and sequence/image are combined and collectively represented
by the stimulus notation $j$. Let $\mu_{jr}$ denote the mean value
over scores for stimulus $j$ and for repetition $r$, i.e. $\mu_{jr}=(\sum_{i}1)^{-1}\sum_{i}u_{ijr}$
. Similarly, $m_{n,jr}$ denotes the $n$-th order central moment
over scores for stimulus $j$ and repetition $r$, i.e. $m_{n,jr}=(\sum_{i}1)^{-1}\sum_{i}(u_{ijr}-\mu_{jr})^{n}$.
Lastly, $\sigma_{jr}$ denotes the sample standard deviation for stimulus
$j$ and repetition $r$, i.e. $\sigma_{jr}=\sqrt{\left((\sum_{i}1)-1\right)^{-1}\sum_{i}(u_{ijr}-\mu_{jr})^{2}}$.
In the previous, the term $\sum_{i}1$ indicates the number of observers
that have offered an opinion score for a given stimulus/repetition,
$jr$. This number of observers could be the same, $I$, or different
per stimulus, if a subjective experiment has been designed in such
a way. The subject rejection procedure in ITU-R BT.500-14 Section
A1-2.3 can be summarized in Algorithm \ref{subjreject}.

\subsection{P.913 Subject Bias Removal}

ITU-T P.913 does not consider repetitions, so the notation $u_{ij}$
denotes the opinion score voted by subject $i$ on stimulus $j$.
The subject bias removal procedure in ITU-T P.913 Section 12.4 can
be summarized in Algorithm \ref{subjbiasrmv}. 

After subject bias removal, we assume that the subject rejection described
in Algorithm \ref{subjreject} is carried out, before calculating
the MOS and the corresponding confidence intervals. Note that P.913
recommends several subject rejection strategies but does not mandate
one (ITU-T P.913 (03/2016) Section 11.4). For simplicity and consistency,
we use the same one as BT.500.

\section{Appendix: First- and Second-Order Partial Derivatives of $L(\theta)$\label{sec:Appendix:partial-derivatives}}

We can derive the first-order and second-order partial derivatives
of $L(\theta)$ with respect to $\psi_{j}$ , $\Delta_{i}$ and $\upsilon_{i}$
as:

\begin{eqnarray*}
\frac{\partial L(\theta)}{\partial\psi_{j}} & = & \sum_{ir}\frac{u_{ijr}-\psi_{j}-\Delta_{i}}{\upsilon_{i}^{2}}\\
\frac{\partial L(\theta)}{\partial\Delta_{i}} & = & \sum_{jr}\frac{u_{ijr}-\psi_{j}-\Delta_{i}}{\upsilon_{i}^{2}}\\
\frac{\partial L(\theta)}{\partial\upsilon_{i}} & = & \sum_{jr}-\frac{1}{\upsilon_{i}}+\frac{(u_{ijr}-\psi_{j}-\Delta_{i})^{2}}{\upsilon_{i}^{3}}\\
\frac{\partial^{2}L(\theta)}{\partial\psi_{j}^{2}} & = & -\sum_{ir}\frac{1}{\upsilon_{i}^{2}}\\
\frac{\partial^{2}L(\theta)}{\partial\Delta_{i}^{2}} & = & -\frac{1}{\upsilon_{i}^{2}}\sum_{jr}1\\
\frac{\partial^{2}L(\theta)}{\partial\upsilon_{i}^{2}} & = & \sum_{jr}\frac{1}{\upsilon_{i}^{2}}-\frac{3(u_{ijr}-\psi_{j}-\Delta_{i})^{2}}{\upsilon_{i}^{4}}
\end{eqnarray*}

\section{An MLE Interpretation of the Plain MOS\label{sec:Appendix:-An-MLE-MOS}}

The plain MOS and its confidence interval can be interpreted using
the notion of maximum likelihood estimation. Consider the model:
\begin{equation}
U_{ijr}=\psi_{j}+\upsilon_{j}X,\label{eq:model_mos}
\end{equation}
where $U_{ijr}$ is the opinion score, $\psi_{j}$ is the true quality
of stimulus $j$ and $\upsilon_{j}$ is the ``ambiguity'' of $j$.
$X\sim N(0,1)$ is i.i.d. Gaussian. Note that this is different from
the proposed model (\ref{eq:model}) where $\upsilon_{i}$ is associated
with the subjects, not the stimuli. We can define the log-likelihood
function for this model as $L(\theta)=\log P(\{u_{ijr}\}|(\{\psi_{j}\},\{\upsilon_{j}\}))$,
and solve for $\{\psi_{j}\}$ and $\{\upsilon_{j}\}$ that maximize
the log-likelihood function, as follows:
\begin{eqnarray*}
\psi_{j} & = & \frac{\sum_{ir}u_{ijr}}{\sum_{ir}1},\\
\upsilon_{j} & = & \sqrt{\frac{\sum_{ir}(u_{ijr}-\psi_{j})^{2}}{\sum_{ir}1}.}
\end{eqnarray*}
 The second-order partial derivative w.r.t. to $\psi_{j}$ is $\frac{\partial^{2}L(\theta)}{\partial\psi_{j}^{2}}=-\frac{1}{\upsilon_{j}^{2}}\sum_{ir}1$.
The 95\% confidence interval of $\psi_{j}$ is then:
\begin{equation}
CI(\psi_{j})=\psi_{j}\pm1.96\frac{\upsilon_{j}}{\sqrt{\sum_{ir}1}}.\label{eq:quality_ci_mos}
\end{equation}
One minor difference between (\ref{eq:quality_ci_mos}) and the 95\%
confidence interval formula in BT.500-14 Section A1-2.2.1 is that
the former uses differential degrees of freedom 0 and the latter uses
1 for the sample standard deviation calculation. In fact, neither
one is fully precise. In the most precise way to calculate the confidence
interval, one should use a Student's \emph{t}-distrubiton with a differential
degrees of freedom 1 (see Section \ref{subsec:Validation-of-Solvers}
and Table \ref{table:synth_ci_coverage} for more discussions).

\begin{algorithm}[t]
\normalsize
\begin{itemize}
\item Input: $u_{ijr}$ for $i=1,...,I$, $j=1,...,J$ and $r=1,...,R$.
\item Initialize $p(i)\leftarrow0$ and $q(i)\leftarrow0$ for $i=1,...,I$.
\item For $j=1,...,J$, $r=1,...,R$:
\begin{itemize}
\item Let $Kurtosis_{jr}=\frac{m_{4,jr}}{\left(m_{2,jr}\right)^{2}}$. 
\item If $2\leq Kurtosis_{jr}\leq4$, then $\epsilon_{jr}=2$; otherwise
$\epsilon_{jr}=\sqrt{20}$.
\item For $i=1,...,I$:
\begin{itemize}
\item If $u_{ijr}\geq\mu_{jr}+\epsilon_{jr}\sigma_{jr}$, then $p(i)\leftarrow p(i)+1.$
\item If $u_{ijr}\leq\mu_{jr}-\epsilon_{jr}\sigma_{jr}$, then $q(i)\leftarrow q(i)+1$.
\end{itemize}
\end{itemize}
\item Initialize $Set_{rej}=\emptyset$.
\item For $i=1,...,I$:
\begin{itemize}
\item If $\frac{p(i)+q(i)}{\sum_{jr}1}\geq0.05$ and $\left|\frac{p(i)-q(i)}{p(i)+q(i)}\right|<0.3$,
then $Set_{rej}\leftarrow Set_{rej}\cup\{i\}.$
\end{itemize}
\item Output: $Set_{rej}$.
\end{itemize}
\caption{ITU-R BT.500 Subject Rejection \cite{bt500}}
\label{subjreject}
\end{algorithm}

\begin{algorithm}[t]
\normalsize
\begin{itemize}
\item Input: 
\begin{itemize}
\item $u_{ij}$ for subject $i=1,...,I$, stimulus $j=1,...,J$.
\end{itemize}
\item For $j=1,...,J$:
\begin{itemize}
\item Estimate MOS of stimulus as $MOS_{j}=(\sum_{i}1)^{-1}\sum_{i}u_{ij}$.
\end{itemize}
\item For $i=1,...,I$:
\begin{itemize}
\item Estimate subject bias as $BIAS_{i}=(\sum_{j}1)^{-1}\sum_{j}(u_{ij}-MOS_{j})$.
\end{itemize}
\item Calculate the subject bias-removed opinion scores $r_{ij}=u_{ij}-BIAS_{i}$,
$i=1,...,I$, $j=1,...,J$.
\item Use $r_{ij}$ instead of $u_{ij}$ as the opinion scores to carry
out the remaining steps.
\end{itemize}
\caption{ITU-T P.913 Subject Bias Removal \cite{p913}}
\label{subjbiasrmv}
\end{algorithm}

%\section*{Acknowledgment}

\ifCLASSOPTIONcaptionsoff
  \newpage
\fi

\end{document}